\documentclass[%
preprint,
 aps,
prb,
]{revtex4-2}

\usepackage{graphicx}
\usepackage{dcolumn}
\usepackage{bm}

\usepackage[top=15mm,bottom=20mm,left=15mm,right=15mm]{geometry}
\usepackage[doublespacing]{setspace}
\usepackage{amsmath}
\usepackage{amssymb}
\usepackage{bm,bbm}
\usepackage{subcaption}
\usepackage{physics,color}

\usepackage{caption}

\begin{document}

\preprint{APS/123-QED}

\title{Quantum effects in rotationally invariant spin glass models}

\author{Yoshinori Hara}
 \email{hara-yoshinori793@g.ecc.u-tokyo.ac.jp}
\affiliation{
Department of Physics, The University of Tokyo, 7-3-1 Hongo, Tokyo 113-0033, Japan}.



\author{Yoshiyuki Kabashima}
\email{kaba@phys.s.u-tokyo.ac.jp}
\affiliation{
 Institute for Physics of Intelligence, 
 The University of Tokyo, 7-3-1 Hongo, Tokyo 113-0033, Japan \\Trans-Scale Quantum Science Institute, The University of Tokyo, 7-3-1 Hongo, Tokyo 113-0033, Japan
}%


\date{\today}

\begin{abstract}
This study investigates the quantum effects in transverse-field Ising spin glass models with rotationally invariant random interactions. 
The primary aim is to evaluate the validity of 
a \textcolor{black}{quasi-static approach} that captures the imaginary-time dependence of the order parameters beyond the conventional static approximation. 
Using the replica method combined with the Suzuki--Trotter decomposition, we 
established a stability condition for the replica symmetric solution, 
which is analogous to the de Almeida--Thouless criterion. 
Numerical analysis of the Sherrington--Kirkpatrick 
model estimates a value of the critical transverse field, $\Gamma_\mathrm{c}$, 
which agrees with previous Monte Carlo-based estimations. 
For the Hopfield model, 
it provides an estimate of $\Gamma_\mathrm{c}$, which has not been previously evaluated.  
For the random orthogonal model, our analysis suggests that quantum effects alter the random first-order transition scenario
in the low-temperature limit.  
This study supports a quasi-static treatment for analyzing quantum spin glasses and may offer useful insights into the analysis of quantum optimization algorithms.
\end{abstract}

\maketitle


\section{Introduction}
Remarkable developments in nanotechnology have stimulated research into the use of quantum mechanics for efficient information processing. Quantum annealing (QA) \cite{kadowaki1998}, also known as adiabatic quantum computation \cite{farhi2000quantumcomputationadiabaticevolution}, is one of the main directions for such efforts. The difficulty of combinatorial optimization stems from the existence of local optima in the objective function, which are separated by high potential barriers. By mapping the objective function onto the Hamiltonian of quantum systems, QA aims to circumvent this difficulty by exploiting quantum tunneling.

This concept sheds new light on the study of quantum spin glasses (QSGs). The performance of QA deteriorates significantly as the energy gap between the ground and first excited states becomes very small. Such situations typically occur during phase transitions. Therefore, there is growing interest in quantum phase transitions exhibited by spin-glass models of the mean-field type, as these serve as analytically soluble models that can characterize the possibilities and limitations of QA \cite{kadowaki1998,doi:10.1143/JPSJ.76.054002,PhysRevB.48.12778,PhysRevLett.101.147204}.

The standard procedure for dealing with QSGs is to employ the Suzuki--Trotter formula (STF) \cite{Trotter1959product,10.1143/PTP.56.1454}. In the case of transverse-field Ising spin models, this procedure transforms each quantum spin operator into $M$ classical spins that interact with each other via ferromagnetic coupling along the imaginary time direction. However, to take $M\to\infty$, which is necessary for retrieving the quantum limit exactly using the STF, many previous studies \cite{bray1980replica,Goldschmidt1990} have ignored this dependence. This treatment is termed the ``static approximation'' (SA) \cite{bray1980replica}. Later, part of the dependence was considered \cite{PhysRevE.96.032112,PhysRevB.109.024431}, which we term the ``\textcolor{black}{quasi-static approach}'' (qSA). However, the qSA ignores the imaginary time dependence of the order parameters, except for those related to the two-point correlation function. To the best of our knowledge, the validity of this treatment has not yet been fully clarified, except in some studies \cite{PhysRevB.41.428} that verified its validity through numerical experiments.

The main goal of this study is to investigate the validity of the qSA. To date, the primary testbed in the study of mean-field QSGs is the quantum Sherrington--Kirkpatrick (SK) model \cite{PhysRevLett.64.2467,PhysRevB.52.384,PhysRevB.41.4858}. However, the examination of a single model alone does not demonstrate the validity of this procedure. Therefore, we consider a family of spin-glass models characterized by rotationally invariant random coupling matrices \cite{PhysRevE.67.046112}, including the SK model as a special case. \par

The remainder of this paper is organized as follows. In Section 2, we focus on the proposed model. In Section 3, we analyze it using the STF and the replica method, maintaining $M$ finite. Depending on the assumed level of replica symmetry breaking (RSB), this yields a set of self-consistent equations defined among the order parameters that are generally imaginary time-dependent. In Section 4, we carefully examine the properties of the solution. 
If the system is replica symmetric, then
we show that the translational invariance along imaginary time and non-negativity of spin correlations always guarantee the existence of a special solution, for which the order parameters, except for those regarding two-point spin correlations, are uniform over imaginary time. This solution corresponds to the qSA. In addition, we derive the local stability condition of the solution against perturbations that break the uniformity of the order parameters. We also show that the condition agrees with that for local stability against the 1-step RSB (1RSB),
which corresponds to the de Almeida--Thouless (AT) condition \cite{de1978stability}.
These results support the use of the qSA in the RS ansatz.  
Section 5 presents the validity of the results obtained by the numerical calculations for the two example systems. 
Additionally, we also analyze another system that exhibited a phase transition of 
another type, called the random first-order transition (RFOT) in the classical case. 
Our analysis suggests that the quantum effects change
the scenario of the RFOT at the low-temperature limit. 
Section 6 presents a summary and discussion.

\section{Model setup}
We consider QSG models defined by the Hamiltonian
\begin{align}
    H = - \sum_{i<j}J_{ij} \hat{\sigma}_i^z \hat{\sigma}_j^z -h_0 \sum_{i=1}^N \hat{\sigma}_i^z -\Gamma \sum_{i=1}^N \hat{\sigma}_i^x, 
    \label{eq:hamiltonian}
\end{align}
where $\hat{\sigma}_i^x$ and $\hat{\sigma}_i^z$ represent Pauli matrices of the $i$th ($i\in \{1,\ldots,N\}$) spin. 
The interaction matrix $J=(J_{ij})$ is constructed as follows:
\begin{align}
    J = ODO^\top 
    \label{eq:RIM}
\end{align}
where $O$ denotes a random $O(N)$ matrix sampled from the Haar measure. 
$D$ is a diagonal matrix whose entries follow a distribution $\rho(\lambda)$. $h_0$ and $\Gamma$ are the vertical and transverse fields, respectively. 

The classical version of this model, which corresponds to the case of $\Gamma = 0$, 
constitute a family of known analytically soluble models 
\cite{Parisi_1995,PhysRevE.67.046112}. 
For instance, the SK model is characterized by
\begin{align}
    \rho(\lambda) = \frac{\sqrt{4-\lambda^2}}{2\pi},
    \label{eq:semi_circ}
\end{align}
whereas the Hopfield model that memorizes $p = \alpha N$ patterns is approximated accurately by 
\begin{align}
    \rho(\lambda) = {\rm max}(1-\alpha,0) \delta(\lambda) + \frac{\sqrt{(\lambda_+ - \lambda)(\lambda -\lambda_-)}}{2\pi \lambda}, 
\end{align}
where ${\rm max}(x,y)$ returns the larger value out of $x$ and $y$, and $\lambda_{\pm} = (1\pm \sqrt{\alpha})^2$.
These models exhibit continuous spin glass phase transitions 
at a sufficiently low temperature. 
By contrast, another type of phase transition, the RFOT \cite{Parisi_1995}, occurs in
\begin{align}
    \rho(\lambda) = \frac{1}{2}\delta(\lambda -J) + \frac{1}{2}\delta(\lambda + J).
    \label{eq:ROM}
\end{align}

Our primary interest is to examine how the quantum effect produced by a nonzero transverse field $\Gamma$
influences the phase transitions in a unified manner. 
\section{Analytical treatment}
\subsection{Suzuki--Trotter decomposition}
We denote the first two terms and the last term of (\ref{eq:hamiltonian}) as $U(\hat{\vb{\sigma}}^z)$ and 
$K(\hat{\vb{\sigma}}^x)$, respectively. Thus, the partition function for (\ref{eq:hamiltonian}) with an inverse temperature 
$\beta > 0$ can be expressed as follows: 
\begin{align}
    Z &= {\rm Tr}e^{-\beta H} = {\rm Tr}e^{-\beta(U+K)} \cr
    &={\rm Tr} \left (e^{-\beta U/M} e^{-\beta K/M} \right )^M
    + O\left (\frac{\beta^2}{M} \right ), 
\end{align}
which is known as the Suzuki--Trotter (ST) decomposition \cite{Trotter1959product,10.1143/PTP.56.1454}. 

The usefulness of this formula comes from the insertion of $M-1$ complete bases into the first term on the right-hand side, which we denote $Z_M$, as
\begin{align}
    Z_M = \sum_{\vb{\sigma}_*^z=\pm 1} 
    \left \langle \vb{\sigma}_1^z \right |
    e^{-\beta U/M} e^{-\beta K/M} \left |\vb{\sigma}_2^z \right \rangle
    \cdots 
    \left \langle \vb{\sigma}_M^z \right |
    e^{-\beta U/M} e^{-\beta K/M} \left |\vb{\sigma}_1^z \right \rangle, 
\end{align}
where $\vb{\sigma}_t^z = (\sigma_{1,t}^z, \cdots, \sigma_{N,t}^z)^\top  
\in \{+1,-1\}^N$, and $\sum_{\sigma^z_\ast=\pm 1} (\cdots)$ stands for the summation over the all possible configurations of $\sigma^z_t$ for $t=1,\ldots,M$. After some algebra, this expression can be rewritten as
\begin{align}
    Z_M = A^{MN} \sum_{\sigma_*^z=\pm 1} 
    \exp\left (
    \sum_{t=1}^M 
    \left [ 
    \frac{\beta U(\sigma_t)}{M} + 
    B \sum_{i=1}^N \sigma_{i,t} \sigma_{i,t+1}
    \right ]
    \right ), 
    \label{eq:STdecomposition}
\end{align}
where we omitted the superscript $z$ from $\sigma^z_t$, denoting it as $\sigma_t$. The constant $A$ and $B$ are defined as 
\begin{align}
   A = \left (\frac{1}{2}\sinh \frac{2 \beta \Gamma}{M} \right )^{1/2}, \quad
    B=\frac{1}{2} \ln \left (\coth \frac{\beta \Gamma}{M} 
    \right ).
\end{align}

Equation (\ref{eq:STdecomposition}) can be interpreted as a partition function for a system consisting of $M$ copies of the classical Ising spin system interacting via ferromagnetic coupling with periodic boundary conditions. This demonstrates that it is possible to study the properties of a quantum system by evaluating the properties of an equivalent classical system and then extrapolating the results to $M\to \infty$. 

\subsection{Replica method}
The partition function for the quantum Hamiltonian can be assessed as 
(\ref{eq:hamiltonian}) by employing the equivalent classical expression
~ (\ref{eq:STdecomposition}). 
However, in this case, we must evaluate the  
the average of $\frac{1}{N} \ln Z_M$ with respect to $J$ to investigate the typical properties of the system under the random generation of $J$. 
Unfortunately, it is technically and computationally difficult to perform this task rigorously. 

To overcome this difficulty, we resort to the replica method, 
which comprises the following two steps:
\begin{enumerate}
    \item First, we assess the moment of (\ref{eq:STdecomposition}), 
    $\mathbb{E}_J \left [Z_M^n \right ]$, for $n =1,2,\ldots \in \mathbb{N}$, where 
    $\mathbb{E}_J [\cdots]$ denotes the average of $\cdots$ with respect to $J$. In practice, this assessment is reduced to a saddle point problem 
    with respect to $n$ copy systems originating from (\ref{eq:STdecomposition}), 
    which are termed ``replicas.'' 
    \item Under an assumption of symmetry with respect to the 
    permutation of indices of the replicas, the saddle point problem yields an
    analytical expression of $\frac{1}{N} \ln \mathbb{E}_J \left [Z_M^n \right ]$ 
    with respect to $n$, which is likely to hold for real numbers $n\in \mathbb{R}$
    as well. We, therefore, analytically continue the expression 
    to $n\in \mathbb{R}$ and assess $\frac{1}{N} \mathbb{E}_J[\ln Z_M]$ 
    using the ``replica trick'' identity
    \begin{align}
        \frac{1}{N}\mathbb{E}_J \left [\ln Z_M \right ] = \lim_{n\to 0}
         \frac{\partial}{\partial n} \frac{1}{N}
        \ln \mathbb{E}_J \left [Z_M^n \right ]. 
        \label{eq:replica_trick}
    \end{align}
\end{enumerate}
The details of each step are as follows. 
\subsubsection{Assessment of $\left [Z_M^n \right ]_J$ for $n \in \mathbb{N}$}
For $n \in \mathbb{N}$, increasing $Z_M$ to a power of $n$ results in 
\begin{align}
    Z_M^{n} &= A^{MNn}
    \sum_{\sigma_{*,*}^*=\pm 1} 
    \exp\left ( \frac{\beta}{2M} 
    \sum_{t,\mu}(\vb{\sigma}_{t}^\mu)^\top J\vb{\sigma}_t^\mu 
    + B \sum_{t,i,\mu} \sigma_{i,t}^\mu \sigma_{i,t+1}^\mu 
    +\frac{\beta h_0}{M} \sum_{t,i,\mu} \sigma_{i,t}^\mu 
    \right ) \cr
    & = A^{MNn}
    \sum_{\sigma_{*,*}^*=\pm 1} 
    \exp\left ( \frac{1}{2} 
    {\rm Tr}\left [J\frac{\beta L}{M} \right ] 
    + B \sum_{t,i,\mu} \sigma_{i,t}^\mu \sigma_{i,t+1}^\mu
    + \frac{\beta h_0}{M} \sum_{t,i,\mu} \sigma_{i,t}^\mu 
    \right ),  
    \label{eq:replicated}
\end{align}
where $\mu=1,\cdots,n$ denotes the replica indices, and 
$L=\sum_{t,\mu} \vb{\sigma}_t^\mu (\vb{\sigma}_t^\mu)^\top$. 
For convenience in assessing $\mathbb{E}_J\left [Z_M^n \right ]$, we introduce the characteristic function of the ensemble of $J$ as
\begin{align}
    G(x) = \mathop{\rm max}_{\Lambda}
    \left ( -\frac{1}{2}\int d\lambda \rho(\lambda) 
    \ln (\Lambda -\lambda)+\frac{\Lambda x}{2} 
    \right )-\frac{1}{2}\ln x - \frac{1}{2}. 
    \label{eq:G_func}
\end{align}

The matrix integral formula
\begin{align}
    \mathbb{E}_J \left [ \exp\left ( \frac{1}{2} 
    {\rm Tr}\left [J\frac{\beta L}{M} \right ]  \right )\right ]
    \simeq \exp \left ( 
    \frac{N}{M} {\rm Tr} G\left ( \frac{\beta L}{N M}\right )
    \right ), 
    \label{eq:E_JZ_Mn}
\end{align}
which holds for $N\gg 1$ when the rank of $L$ is $o(N)$, plays a key role \cite{Parisi_1995}. 
Because $\frac{L}{N} = \frac{1}{N}\sum_{t,\mu} \vb{\sigma}_t^\mu 
(\vb{\sigma}_{t}^\mu )^\top \in \mathbb{R}^{N \times N}$ and 
$Q =(q_{t,s}^{\mu,\nu}) = 
\left ( \frac{1}{N}\vb{\sigma}_t^\mu \cdot \vb{\sigma}_s^\nu \right)
\in \mathbb{R}^{nM \times nM}$ share all non-zero eigenvalues and 
$G(0) = 0$, 
we can rewrite (\ref{eq:E_JZ_Mn}) as
\begin{align}
    \mathbb{E}_J \left [ \exp\left ( \frac{1}{2} 
    {\rm Tr}\left [J\frac{\beta L}{M} \right ]  \right )\right ]
    \simeq \exp \left ( 
    N {\rm Tr} G\left ( \frac{\beta Q}{M}\right )
    \right ). 
    \label{eq:E_JZ_Mn2}
\end{align}
By inserting this into the computing of 
$\mathbb{E}_J\left [Z_M^n \right]$ together with 
trivial identities 
\begin{align}
    1 &= N \int {\mathrm d}q_{t,s}^{\mu,\nu}
    \delta \left (\vb{\sigma}_t^\mu \cdot \vb{\sigma}_s^\nu 
    -N q_{t,s}^{\mu,\nu} \right ) \cr
    & = \frac{N}{4\pi M^2} \int {\mathrm d} q_{t,s}^{\mu, \nu}
    \int_{-i \infty}^{+i \infty}{\mathrm d}\tilde{q}_{t,s}^{\mu,\nu}
    \exp \left ( \frac{\tilde{q}_{t,s}^{\mu,\nu} }{2M^2}
    \left (\sum_{i=1}^N \sigma_{i,t}^\mu \sigma_{i,s}^\nu -Nq_{t,s}^{\mu, \nu} \right ) 
    \right )
\end{align}
for $t,s \in \{1,\cdots,M\}$ and $\mu,\nu\in \{1,\cdots,n\}$,  
which decouples the dependence on site indices $i \in \{1,\ldots, N\}$, yields
an expression for the saddle point assessment for $N\gg 1$ as
\begin{align}
    \frac{1}{N} \ln \mathbb{E}_J \left [Z_M^n \right ]
    \simeq \mathop{\rm Extr}
    \left \{ {\rm Tr} G\left (\frac{\beta Q}{M} \right )
    -\frac{1}{2M^2}\sum_{t,s,\mu,\nu} \tilde{q}_{t,s}^{\mu,\nu} q_{t,s}^{\mu,\nu}
    + \ln {\rm Tr} \exp \left (-H_{\rm eff} \right )
    \right \},  
    \label{eq:free_energy_for_finite_n}
\end{align}
where 
\begin{align}
    H_{\rm eff} = -B\sum_{t,\mu}\sigma_t^\mu \sigma_{t+1}^\mu 
    -\frac{\beta h_0}{M} \sum_{t,\mu} \sigma_{t}^\mu 
    - \frac{1}{2M^2}\sum_{t,s,\mu,\nu} \tilde{q}_{t,s}^{\mu,\nu}
    \sigma_t^\mu\sigma_s^\nu, 
    \label{eq:effective_hamiltonian}
\end{align}
$\mathop{\rm Extr}\{\cdots\}$ denotes the operation of extremization 
with respect to $q_{t,s}^{\mu,\nu}$ and $\tilde{q}_{t,s}^{\mu,\nu}$. 
This implies that for $n\in \mathbb{N}$, one can evaluate the properties of 
the system by solving the saddle-point equations 
\begin{align}
    \tilde{q}_{t,s}^{\mu,\nu} &= 
    2M^2 \frac{\partial }{\partial q_{t,s}^{\mu,\nu}} {\rm Tr} G\left (\frac{\beta Q}{M} \right ),
     \label{eq:saddle_point_eq_general1}\\
    q_{t,s}^{\mu,\nu} &= \left \langle 
    \sigma_t^\mu \sigma_s^\nu \right \rangle_{\rm eff},
    \label{eq:saddle_point_eq_general2}
\end{align}
where $\left \langle \cdots \right \rangle_{\rm eff}$
indicates the average with respect to the effective Boltzmann distribution
$p_{\rm eff}(\vb{\sigma}) \propto \exp\left (-H_{\rm eff} \right )$  
for $\vb{\sigma}=(\sigma_t^\mu) \in \{+1,-1\}^{nM}$.\par


\subsubsection{Replica symmetry (RS) and analytic continuation}
For $n\in \mathbb{N}$, we evaluate 
(\ref{eq:free_energy_for_finite_n}) 
exactly by solving (\ref{eq:saddle_point_eq_general1}) and (\ref{eq:saddle_point_eq_general2}). Unfortunately, we cannot use the resultant expression 
directly 
for computing (\ref{eq:replica_trick}). 
Focusing on the following property is the key to resolving this issue. 
Equation (\ref{eq:replicated}) is invariant under any permutation among the 
replica indices $\mu\in \{1,\ldots, n\}$, which is termed the ``replica symmetry''. 
Therefore, it is natural to assume that the solutions to (\ref{eq:saddle_point_eq_general1}) and (\ref{eq:saddle_point_eq_general2})
exhibit the same property. This limits the solution to the form of
\begin{align}
    \tilde{q}_{t,s}^{\mu, \nu} &= 
    \left \{
    \begin{array}{rl}
    \tilde{\chi}_{t,s} + \tilde{q}_{t,s}, & (\mu = \nu) \\
    \tilde{q}_{t,s}, & (\mu \ne \nu)
    \end{array}
    \right .  , 
    \label{eq:RS1} \\
    {q}_{t,s}^{\mu, \nu} &= 
    \left \{
    \begin{array}{rl}
    1, & (\mu = \nu, t=s) \\
    {\chi}_{t,s} + {q}_{t,s}, & (\mu = \nu, t \ne s) \\
    {q}_{t,s}, & (\mu \ne \nu, t \ne s)
    \end{array}
    \right .  
    \label{eq:RS2}
\end{align}
which is called the replica symmetric (RS) solution. 
Here, $\tilde{\chi}_{t,s}$ and $\chi_{t,s}$ denote the order parameters 
for the two-point correlation function. 

In addition, the periodic boundary conditions with respect to the indices of Trotter slices
$t,s$ make $\chi_{t,s}$, $q_{t,s}$, $\tilde{\chi}_{t,s}$, and $\tilde{q}_{t,s}$
symmetric circulant matrices
because $\chi_{t,s} = \chi(t-s) = \chi(s-t)$, 
$q_{t,s} = q(t-s) = q(s-t)$, 
$\tilde{q}_{t,s} = \tilde{q}(s-t)= \tilde{q}(t-s)$, and 
$\tilde{\chi}_{t,s} = \tilde{\chi}(s-t)= \tilde{\chi}(t-s)$ \cite{doi:10.7566/JPSJ.94.044005}. 
This implies that these matrices are commonly diagonalized 
using an orthonormal basis ${\mathcal U} = ({\mathcal U}_{t,j})$, where 
\begin{align}
\label{eq:odd_U}
    {\mathcal U}_{t,j} = 
    \left \{
    \begin{array}{cl}
    \frac{1}{\sqrt{M}} & (j=0) \\
    \sqrt{\frac{2}{M}} \cos \left ( \frac{\pi (j+1) t}{M} \right )& (j\in \{1,3,..,M-2\}) \\
    \sqrt{\frac{2}{M}} \sin \left ( \frac{\pi j t}{M} \right ) & (j\in \{2,4,...,M-1\}) \\
    \end{array}
    \right .
\end{align}
for odd $M$ and 
\begin{align}
\label{eq:even_U}
    {\mathcal U}_{t,j} = 
    \left \{
    \begin{array}{cl}
    \frac{1}{\sqrt{M}} & (j=0) \\
    \sqrt{\frac{2}{M}} \cos \left ( \frac{\pi (j+1) t}{M} \right )& (j\in \{1,3,..,M-3\}) \\
    \sqrt{\frac{2}{M}}\sin \left ( \frac{\pi j t}{M} \right ) & (j\in \{2,4,...,M-2\}) \\
    \frac{(-1)^t}{\sqrt{M}} & (j = M-1)
    \end{array}
    \right .
\end{align}
for even $M$. 
Using this basis, the eigenvalues of $\chi_{t,s}$ and $q_{t,s}$ are 
expressed as
\begin{align}
\eta_j = \sum_{t,s}{\mathcal U}_{t,j} {\mathcal U}_{s,j} \chi(t-s)
\end{align}
and 
\begin{align}
    r_j = \sum_{t,s}{\mathcal U}_{t,j} {\mathcal U}_{s,j} q(t-s), 
\end{align}
respectively, and similarly for $\tilde{\chi}_{t,s}$ and $\tilde{q}_{t,s}$, 
which offers the eigenvalues of $Q$ as
\begin{align}
    w_{0,j} &=  \eta_j + nr_j,  
    \label{eq:eigenvalueQ1}\\
    w_{1,j} &= \eta_j  ,  
    \label{eq:eigenvalueQ2}
\end{align}
where $j \in \{0,\cdots, M-1\}$ and $w_{1,j}$ are degenerated by $n-1$ for each $j$. 
Inserting these into (\ref{eq:free_energy_for_finite_n}) in conjunction with 
an identity 
\begin{align}
    \exp \left (\frac{1}{2M^2} \sum_{t,s, \mu, \nu}
    \tilde{q}(t-s) \sigma_t^\mu \sigma_s^\nu \right )
    =\int {\mathrm D}^M \vb{z} 
    \exp \left (\sum_{t,\mu}\frac{1}{M} \left (\sqrt{\tilde{Q}}\vb{z} \right )_t \sigma_t^\mu
    \right ), 
\end{align}
where ${\mathrm D}^M\vb{z} = (\sqrt{2\pi})^{-M/2}\exp\left (-\sum_{t}z_t^2/2 \right ) \prod_{t=0}^{M-1} \dd{z_t}$
and $\sqrt{\tilde{Q}}$ denotes the Cholesky decomposition of $\tilde{Q} = (\tilde{q}(t-s))$, 
yields an expression
\begin{align}
     & \frac{1}{N} \ln \mathbb{E}_J \left [Z_M^n \right ] \cr
     & \simeq \mathop{\rm Extr}
    \left \{  \sum_{j=0}^{M-1} 
    \left (G\left (\frac{\beta (\eta_j + nr_j) }{M} \right )
    +(n-1)G\left (\frac{\beta \eta_j }{M} \right ) \right ) 
    + \ln \Xi_{\rm RS}(n) \right . \cr
    & 
    \hspace{1.5cm} -\frac{n}{2M^2}\sum_{t,s} (\tilde{\chi}(t-s)+\tilde{q}(t-s) )
    (\chi(t-s)+q(t-s)) \cr
    & \hspace{1.5cm} \left . -\frac{n(n-1)}{2M^2} 
    \sum_{t,s} \tilde{q}(t-s) q(t-s) \right \}   
\end{align}
where 
\begin{align}
   \Xi_{\rm RS} (n) &= 
   \sum_{\sigma_*^* = \pm 1} 
   \exp \left (B\sum_{t,\mu}\sigma_t^\mu \sigma_{t+1}^\mu 
   + \frac{\beta h_0}{M}\sum_{t,\mu}\sigma_t^\mu + \frac{1}{2M^2}\sum_{t,s,\mu}
   \tilde{\chi}(t-s)\sigma_t^\mu \sigma_s^\mu \right . \cr
   &\hspace{1cm} \left . +  \frac{1}{2M^2}\sum_{t,s,\mu, \nu}
   \tilde{q}(t-s)\sigma_t^\mu \sigma_s^\nu
   \right ) \cr
   &=\int {\mathrm D}\vb{z}
   \left (
   \sum_{\sigma_* = \pm 1} 
   \exp \left (B\sum_{t}\sigma_t \sigma_{t+1} 
   + \frac{1}{M}\sum_{t}\left (\beta h_0 + \left (\sqrt{\tilde{Q}} \vb{z} \right )_t 
   \right )\sigma_t \right ) \right ) \cr
    &  \hspace{1cm} \left . \left .
    + \frac{1}{2M^2}\sum_{t,s}
    \tilde{\chi}(t-s)\sigma_t \sigma_s \right ) \right )^n,  
\end{align}
which can also be defined for $n \in \mathbb{R}$. 
Therefore, we use these expressions to compute (\ref{eq:replica_trick}), 
which leads to
\begin{align}
 & \frac{1}{N} \mathbb{E}_J \left [ \ln Z_M^n \right ] \simeq \mathop{\rm Extr}
    \left \{  \sum_{j=0}^{M-1} 
    \left (G\left (\frac{\beta \eta_j }{M} \right )
    +\frac{\beta r_j}{M} G^\prime \left (\frac{\beta \eta_j }{M} \right ) \right ) \right . \cr
    &\hspace{0.3cm}  
    -\frac{1}{2M^2}\sum_{t,s} \left (\tilde{\chi}(t-s) (\chi(t-s) + {q}(t-s)) 
    + \tilde{q}(t-s) \chi(t-s) \right ) \cr
    & \left . \hspace{0.3cm}
    + \int {\mathrm D}^M \vb{z} \ln {\rm Tr} \exp \left (-H_{\rm eff}^{\rm RS} \right )
    \right \},  
    \label{eq:RS_free_energy}
\end{align}
where 
\begin{align}
    H_{\rm eff}^{\rm RS}
    = & -B\sum_{t}\sigma_t \sigma_{t+1} 
   - \frac{1}{M} \sum_{t}\left (\beta h_0 + \left (\sqrt{\tilde{Q}} \vb{z} \right )_t 
   \right )\sigma_t  - \frac{1}{2M^2}\sum_{t,s}
    \tilde{\chi}(t-s)\sigma_t \sigma_s. 
    \label{eq:RS_effective}
\end{align}
The solution to the saddle point problem in (\ref{eq:RS_free_energy}) is obtained by 
solving 
\begin{align}
    & \tilde{\chi}(t-s) = 2 M\beta \sum_{j} \mathcal{U}_{t,j}\mathcal{U}_{s,j}
    G^\prime \left ( \frac{\beta \eta_j}{M} \right ),  
    \label{eq:RSsp1}\\
    & \tilde{q}(t-s) = 2 \beta^2 \sum_{j} \mathcal{U}_{t,j}\mathcal{U}_{s,j} 
    r_j G^{\prime \prime}\left ( \frac{\beta \eta_j}{M} \right ),  
    \label{eq:RSsp2}\\
    & \chi (t-s) = \int {\mathrm D}^M\vb{z} \left ( 
    \left \langle \sigma_t \sigma_s \right \rangle_{\rm RS} 
    -\left \langle \sigma_t  \right \rangle_{\rm RS} 
    \left \langle \sigma_s \right \rangle_{\rm RS} 
    \right ),  
    \label{eq:RSsp3}\\
    & q(t-s) = \int {\mathrm D}^M \vb{z} 
    \left \langle \sigma_t  \right \rangle_{\rm RS} 
    \left \langle \sigma_s \right \rangle_{\rm RS}, 
    \label{eq:RSsp4}
\end{align}
where $\left \langle \cdots \right \rangle_{\rm RS}$ denotes the average of $\cdots$ with respect to 
the Boltzmann distribution for (\ref{eq:RS_effective}). 

Here, we discuss the relationship between the aforementioned and earlier results. 
For the SK model, where $G(x)=x^2/4$, 
(\ref{eq:RS_free_energy}) is reduced to the result of $p=2$ in 
\cite{Goldschmidt1990}. However, \cite{Goldschmidt1990} further assumed that all order parameters are uniform in imaginary time, which we term the static approximation (SA) \cite{bray1980replica}, to analytically extrapolate the expression to $M\to\infty$. 
Employing this approximation in our formulation yields 
\begin{align}
     \frac{1}{N} \mathbb{E}_J \left [ \ln Z_M \right ] \simeq \mathop{\rm Extr} \qty[G(\beta \chi)+\beta qG'(\beta \chi)-\frac{1}{2}\qty[(\tilde{\chi}+\tilde{q})(\chi+q)-\tilde{q}q]+\int\mathrm{D}z_0\,\ln \int\mathrm{D}y_0 \mathrm{Tr}\, \exp\qty(-H^\mathrm{SA}_\mathrm{eff})]
    \label{eq:SA_free_energy}, 
\end{align}
where $\chi$, $q$, $\tilde{\chi}$, and $\tilde{q}$ are scalars,  
\begin{align}
    H^\mathrm{SA}_\mathrm{eff} = -B\sum_{t}\sigma_t\sigma_{t+1} - \frac{1}{M}\qty(\beta h_0+\sqrt{\tilde{\chi}}y_0)\sum_{t}\sigma_t,  
    \label{eq:H_eff_SA}
\end{align}
and ${\mathrm D}x = (2\pi)^{-1/2} \dd{x} \exp\left (-x^2/2 \right )$. 
This expression allows us to obtain the quantum limit $M\to \infty$ analytically. 
More precisely, employing the STF 
to the integrant of the last term of (\ref{eq:SA_free_energy}) in the reverse direction yields 
\begin{align*}
    \lim_{M\to \infty} \int\mathrm{D}y_0 \mathrm{Tr}\, \exp\qty(-H^\mathrm{SA}_\mathrm{eff})&=\int\mathrm{D}y_0\,\mathrm{Tr}\,\exp\qty(\beta\qty\Big[\Gamma\hat{\sigma}^x+\frac{1}{\beta}\left (\sqrt{\tilde{\chi}}y_0+\beta h_0 \right )\hat\sigma^z])\\
    &=\int\mathrm{D}y_0\,2\cosh\qty(\beta\sqrt{\Gamma^2+\frac{\left (\sqrt{\tilde{\chi}}y_0+\beta h_0 \right )^2}{\beta^2}}). 
\end{align*}
Further, this enables us to take $\beta \to \infty$ analytically. 
For $h_0=0$, this provides the critical value of the transverse field 
between the paramagnetic and spin glass phases at the vanishing temperature $\Gamma_{\rm c}$ as the solution of 
\begin{align*}
    1=2 xG''(x), \,\,\,\mathrm{for}\,\,x\,\,\,\mathrm{s.t.}\,\,x=\frac{1}{\Gamma_{\rm c}-2G'(x)}, 
\end{align*}
which gives $\Gamma_{\rm c}=2$ for the SK model and $\Gamma_{\rm c}=3+2\sqrt{3}$ 
for the Hopfield model with $\alpha=2$.

The SA enables analytical treatment at low temperatures by analytically determining the quantum limit $M\to \infty$. 
However, the SA solution does not satisfy the saddle-point conditions 
(\ref{eq:RSsp1})--(\ref{eq:RSsp4}). 
This is due to the presence of 
the one-dimensional couplings $-B\sum_{i} \sigma_i \sigma_{i+1}$ in (\ref{eq:H_eff_SA}) 
creates an imaginary time dependence for the two-point correlation function $\chi(t-s)$
in (\ref{eq:RSsp3})
even if its conjugate $\tilde{\chi}$ is uniform over imaginary time. 
This implies that we must consider the imaginary time dependence 
of $\tilde{\chi}$ and $\chi$ to construct solutions satisfying 
the saddle-point conditions in (\ref{eq:RSsp1})--(\ref{eq:RSsp4}).

\section{Quasi-static solution and its stability}
\subsection{Quasi-static solution}
Although maintaining the imaginary time dependence of $\tilde{\chi}$ and $\chi$ is essential, 
we can construct the solutions to (\ref{eq:RSsp1})--(\ref{eq:RSsp4}) under the assumption that 
the other order parameters are uniform over imaginary time. 
We term this type of treatment ``quasi-static.'' 

To demonstrate this, we evaluate $\tilde{r}_j := \sum_{t,s}{\mathcal U}_{t,j}{\mathcal U}_{s,j}\tilde{q}(t-s)$
and ${r}_j = \sum_{t,s}{\mathcal U}_{t,j}{\mathcal U}_{s,j}{q}(t-s)$ using (\ref{eq:RSsp2}) and (\ref{eq:RSsp4}). 
Exploiting the orthogonality of ${\mathcal U}_{t,j}$, we obtain
\begin{align}
    \tilde{r}_j &= 2\beta^2 r_j G^{\prime\prime} \left (\frac{\beta \eta_j}{M} \right ), \\
    r_j &=  \sum_{t,s} {\mathcal U}_{t,j}{\mathcal U}_{s,j}
    \int {\mathrm D}^M \vb{z} \left \langle \sigma_t \right \rangle_{\rm RS}\left \langle \sigma_s \right \rangle_{\rm RS}.
\end{align}
At this point, we assume $\tilde{r}_j =0 $ except for $j = 0$, which implies that 
$\tilde{q}(t-s)$ is uniform over imaginary time, that is, $\tilde{q}(t-s) = \tilde{q}$. 
This indicates that, in (\ref{eq:RS_effective}),
\begin{align}
    \sqrt{\tilde{Q}} \vb{z} = \sqrt{M} {\mathcal U} \times
    \left (
    \begin{array}{cccc}
    \sqrt{\tilde{q}} & 0 & \cdots & 0 \\
    0 & 0 & \cdots & 0 \\
    \vdots & \vdots & \ddots & \vdots \\
    0 & 0 & \cdots & 0 
    \end{array}
    \right ) 
    \left (
    \begin{array}{c}
    z_0(=z) \cr
    z_1 \cr
    \vdots \cr
    z_{M-1} 
    \end{array}
    \right )
    = \left (
    \begin{array}{c}
    \sqrt{\tilde{q}} z \cr
    \sqrt{\tilde{q}} z \cr
    \vdots \cr
    \sqrt{\tilde{q}} z 
    \end{array}
     \right )
\end{align}
holds. 
This makes $\left \langle \sigma_t \right \rangle_{\rm RS}$ uniform over imaginary time
$t$ for an arbitrary $\vb{z} \in \mathbb{R}^M$ as well, which self-consistently guarantees $r_j =0 $ except for $j \ne 0$.

In practice, a quasi-static solution can be obtained by solving 
\begin{align}
    &\tilde{q} = 2\beta^2 q G^{\prime\prime} \left ( \frac{\beta }{M}\sum_{s,t} \chi(t-s) \right ), \\
    &q = \frac{1}{M} \sum_{t} \int {\mathrm D}z \left \langle \sigma_t \right \rangle_{\rm RS}^2,  
\end{align}
for an effective Hamiltonian with random, but uniform, effective fields
\begin{align}
    H_{\rm eff}^{\rm RS(qSA)}
    = & -B\sum_{t}\sigma_t \sigma_{t+1} 
   - \frac{1}{M}\left (\beta h_0 + \sqrt{\tilde{q}} z   \right )\sum_{t} \sigma_t  - \frac{1}{2M^2}\sum_{t,s}
    \tilde{\chi}(t-s)\sigma_t \sigma_s, 
    \label{eq:RS_qSA_effective}
\end{align}
together with the conditions (\ref{eq:RSsp1}) and (\ref{eq:RSsp3}).

\subsection{Stability analysis}

\textcolor{black}{The quasi-static solution, which 
physically means that a uniformity 
\begin{align}
\label{eq:uniform_q}
 q_{t,s}^{\mu, \nu} = N^{-1} \sum_{i=1}^N\mathbb{E}_J 
\left [ 
\left \langle \sigma_{i,t}^\mu \right \rangle 
\left \langle \sigma_{i,s}^\nu \right \rangle 
\right ] =  n^{-1} \sum_{\rho=1}^n M^{-1} \sum_{u=1}^M N^{-1} \sum_{i=1}^N
 \mathbb{E}_J 
\left [ 
\left \langle \sigma_{i,u}^\rho \right \rangle^2 
\right ]
\end{align}
holds for $\mu\neq\nu$ in the computation of (\ref{eq:free_energy_for_finite_n}), 
is a special solution that satisfies (\ref{eq:RSsp1})--(\ref{eq:RSsp4}).
Its existence as a mathematical solution, however, does not necessarily imply that it is physically realized.
The reason why such a solution appears is that the replicated partition function (\ref{eq:STdecomposition}) possesses a symmetry that remains invariant under shifts of the imaginary time variable
$t$, which, under discretization, is equivalently expressed as a rotational symmetry on the imaginary-time circle.
In finite systems, whenever such a symmetry is present, states that respect it -- namely, the quasi-static solution in the present case -- are guaranteed to occur.
In contrast, as exemplified by the spontaneous breaking of the 
$Z_2$ symmetry in the ferromagnetic Ising model without an external field and the replica-symmetry breaking in the SK model, physically realized states in infinite systems may spontaneously break the symmetry inherent in the partition or replicated partition function.
More concretely, modes with $j\ne 0$ in (\ref{eq:odd_U}) or (\ref{eq:even_U}) 
may break the uniformity of (\ref{eq:uniform_q}) in the present system. 
Therefore, it is essential to investigate the thermodynamic stability of the quasi-static solution.}

\textcolor{black}{For this purpose, }
we examined the local stability against 
perturbations that break the imaginary time uniformity. 
In the first place, we set 
\begin{align}
    \tilde{Q} = \tilde{q} \vb{1}_{M\times M} + \Delta \tilde{Q}
\end{align}
in (\ref{eq:RS_effective}), where $\vb{1}_{M\times M}$ denotes $M\times M$ matrix whose entries are all ones and 
$\Delta \tilde{Q}=(\Delta \tilde{q}(t-s))$ is the $M\times M$ perturbation matrix that is symmetric and circulant. 
Linearizing (\ref{eq:RSsp2}) and (\ref{eq:RSsp4}) with respect to 
$\Delta \tilde{Q}$ around the quasi-static solution yields
\begin{align}
    \Delta \tilde{q}(t-s) &= 2 \beta^2 \sum_{j} {\mathcal U}_{t,j}{\mathcal U}_{s,j}\sum_{t^\prime, s^\prime}
    {\mathcal U}_{t^\prime,j}{\mathcal U}_{s^\prime,j}
    \Delta q(t-s) G^{\prime\prime} \left (\frac{\beta \eta_j}{M} \right ), \label{eq:Linearlization1}\\
    \Delta q(t-s) &= \int {\mathrm D}z \sum_{t^\prime, s^\prime}
    (\left \langle \sigma_t \sigma_{t^\prime} \right \rangle_{\rm RS} 
    -\left \langle \sigma_t \right \rangle_{\rm RS} \left \langle \sigma_{t^\prime} \right \rangle_{\rm RS})
    \Delta \tilde{q}(t^\prime-s^\prime)
    (\left \langle \sigma_{s^\prime} \sigma_s \right \rangle_{\rm RS} 
    -\left \langle \sigma_{s^\prime} \right \rangle_{\rm RS} \left \langle \sigma_s \right \rangle_{\rm RS}). 
    \label{eq:Linearlization2}
\end{align}
Here, (\ref{eq:RS_qSA_effective}) guarantees that the matrix
$C(z):=(\left \langle \sigma_t \sigma_{s} \right \rangle_{\rm RS} 
    -\left \langle \sigma_t \right \rangle_{\rm RS} \left \langle \sigma_{s} \right \rangle_{\rm RS})$
is symmetric and circulant for an arbitrary random number $z$.     
This indicates that the matrices $C(z)$ and $\Delta\tilde{Q}$ can be simultaneously
diagonalized using the identical basis ${\mathcal U}$. 
By employing diagonalization, (\ref{eq:Linearlization1}) and (\ref{eq:Linearlization2}) are converted into the following expression:
\begin{align}
    \Delta \tilde{r}_j = {2 \beta^2}G^{\prime \prime} \left ( \frac{\beta \eta_j}{M} \right )
    \int {\mathrm D}z 
    \left [T_j(z) \right ]^2 
     \times \Delta \tilde{r}_j, 
\end{align}
where $j\in \{0,\ldots, M-1\}$, $\Delta \tilde{r}_j :=\sum_{t,s} {\mathcal U}_{t,j}{\mathcal U}_{s,j} \Delta \tilde{q}(t-s)$, 
and $T_j(z):= 
    \sum_{t,s}
    {\mathcal U}_{t,j}{\mathcal U}_{s,j}
    (\left \langle \sigma_t \sigma_{s} \right \rangle_{\rm RS} 
    -\left \langle \sigma_t \right \rangle_{\rm RS} \left \langle \sigma_{s} \right \rangle_{\rm RS})$
represents the $j$th eigenvalue of $C(z)$. 

The positive definiteness of the covariance matrix $C(z)$ guarantees that $T_j(z) >0$ 
for $\forall{j} \in \{0,\ldots,M-1\}$. 
In addition, (\ref{eq:G_func}) indicates 
\begin{align}
    G^{\prime\prime}(x) = \frac{ \int d\lambda \rho(\lambda)(\Lambda(x) -\lambda)^{-2} 
    -\left ( \int d\lambda \rho(\lambda)(\Lambda(x) -\lambda)^{-1} \right )^2}
    {\int d\lambda \rho(\lambda)(\Lambda(x) -\lambda)^{-2} 
    \left ( \int d\lambda \rho(\lambda)(\Lambda(x) -\lambda)^{-1} \right )^2}
     > 0, 
\end{align}
where $\Lambda(x)$ is the solution to 
$\int d\lambda \rho(\lambda)(\Lambda -\lambda)^{-1} = x$. 
This means that the local stability condition of the RS quasi-static solution is 
given by 
\begin{align}
    \mathop{\rm max}_{j\in \{0,\ldots,M-1\}} 
    \left \{2\beta^2 G^{\prime\prime}\left ( \frac{\beta \eta_j}{M} \right ) 
    \int {\mathrm D}z \left [ T_j(z) \right ]^2 
    \right \} < 1. 
    \label{eq:general_stability}
\end{align}

If $G^{\prime \prime}(x)$ does not decrease for $x>0$, 
which we assume hereafter, and holds for the SK and Hopfield models, 
the expression of (\ref{eq:general_stability})
is further simplified. 
The nature of the one-dimensional interactions with 
positive coupling constant $B$ makes all entries of $C(z)$ positive. 
Thus, the Perro--Frobenius theorem ensures that 
the largest eigenvalue of this matrix is 
$\Lambda_0(z) = \sum_{t,s} {\mathcal U}_{t,0} {\mathcal U}_{s,0} 
(\left \langle \sigma_t \sigma_{s} \right \rangle_{\rm RS} 
    -\left \langle \sigma_t \right \rangle_{\rm RS} \left \langle \sigma_{s} \right \rangle_{\rm RS})
=
M^{-1}\sum_{t,s} (\left \langle \sigma_t \sigma_{s} \right \rangle_{\rm RS} 
    -\left \langle \sigma_t \right \rangle_{\rm RS} \left \langle \sigma_{s} \right \rangle_{\rm RS})$. 
Applying a similar argument to the matrix $(\chi(t-s))$ indicates that $\mathop{\rm max}_{j}\{\eta_j\} = \eta_0
=M^{-1} \sum_{t,s} \chi(t-s)$, which  
implies that $\mathop{\rm max}_{j}
\{G^{\prime\prime}(\beta \eta_j/M) \} = G^{\prime\prime}(\beta \eta_0/M)=G^{\prime\prime}\left ({\beta}M^{-1} 
\sum_{t,s} \chi(t-s) \right )$. 
Combining these results indicates that (\ref{eq:general_stability}) is simplified as follows: 
\begin{align}
    2\beta^2G^{\prime\prime}\left (\frac{\beta}{M} \sum_{t,s} \chi(t-s) \right )
    \int {\mathrm D}z \left  [ \frac{1}{M}\sum_{t,s} (\left \langle \sigma_t \sigma_{s} \right \rangle_{\rm RS} 
    -\left \langle \sigma_t \right \rangle_{\rm RS} \left \langle \sigma_{s} \right \rangle_{\rm RS})
    \right ]^2 < 1. 
    \label{eq:AT_RS}
\end{align}

Equations (\ref{eq:general_stability}) and (\ref{eq:AT_RS}) have a different meaning. 
the saddle point condition under the 1-step replica symmetry breaking (1RSB)
ansatz is given by 
\begin{align}
    \tilde{\chi}(t-s) &= 2M\beta \sum_{j}{\mathcal U}_{t,j}{\mathcal U}_{s,j}G^\prime
    \left (\frac{\beta \eta_j}{M} \right ), \label{eq:1RSB1} \\
    \tilde{q}^1(t-s)- \tilde{q}^0(t-s) 
    &= \frac{2\beta M}{m} \sum_{j}{\mathcal U}_{t,j}{\mathcal U}_{s,j}
    \left [
    G^\prime\left( \frac{\beta}{M} [\eta_j + m(r_j^1-r_j^0)] \right )
    -G^\prime\left( \frac{\beta \eta_j}{M}\right ) \right ], \label{eq:1RSB2} \\
    \tilde{q}^0(t-s) &= 2 \beta^2 \sum_{j} {\mathcal U}_{t,j}{\mathcal U}_{s,j} r_j^0 G^{\prime \prime}
    \left (\frac{\beta}{M}\left [\eta_j + m(r_j^1-r_j^0) \right ] \right ), \label{eq:1RSB3} \\
    \chi(t-s) &= \int {\mathrm D}^M\vb{z}
    \frac{\int {\mathrm D}^M \vb{y} Z_{\rm 1RSB}(\vb{h})^m 
    \left ( \left \langle \sigma_t \sigma_s \right \rangle_{\rm 1RSB}
    -\left \langle \sigma_t \right \rangle_{\rm 1RSB} \left \langle \sigma_s \right \rangle_{\rm 1RSB} \right )}
    {\int {\mathrm D}^M \vb{y} Z_{\rm 1RSB}(\vb{h})^m},
    \label{eq:1RSB4} \\
    q^1(t-s) &= \int {\mathrm D}^M\vb{z}
    \frac{\int {\mathrm D}^M \vb{y} Z_{\rm 1RSB}(\vb{h})^m 
    \left \langle \sigma_t \right \rangle_{\rm 1RSB} \left \langle \sigma_s \right \rangle_{\rm 1RSB} }
    {\int {\mathrm D}^M \vb{y} Z_{\rm 1RSB}(\vb{h})^m}.  \label{eq:1RSB5}\\
    q^0(t-s) &= 
    \int {\mathrm D}^M\vb{z}
    \left (
    \frac{\int {\mathrm D}^M \vb{y} Z_{\rm 1RSB}(\vb{h})^m \left \langle \sigma_t \right \rangle_{\rm 1RSB}}
    {\int {\mathrm D}^M \vb{y} Z_{\rm 1RSB}(\vb{h})^m}
    \right )
    \left (
    \frac{\int {\mathrm D}^M \vb{y} Z_{\rm 1RSB}(\vb{h})^m \left \langle \sigma_s \right \rangle_{\rm 1RSB}}
    {\int {\mathrm D}^M \vb{y} Z_{\rm 1RSB}(\vb{h})^m}\right ), \label{eq:1RSB6}   
\end{align}
where $m$ is Parisi's breaking parameter. 
$r_j^l =\sum_{t,s} {\mathcal U}_{t,j}{\mathcal U}_{s,j} q^l(t-s)$ ($l \in \{0,1\}$), 
and $\vb{h} =\sqrt{\tilde{Q}^1 -\tilde{Q}^0} \vb{y} + \sqrt{\tilde{Q}^0} \vb{z} \in \mathbb{R}^M$, 
where $\tilde{Q}^l = (\tilde{q}^l(t-s))$ ($l \in \{0,1\}$). 
For $H_{{\rm eff}}^{\rm 1RSB}= -B \sum_{t}\sigma_t \sigma_{t+1} -M^{-1} \sum_{t} (\beta h_0 + h_t)\sigma_t 
-(2M^2)^{-1} \sum_{t,s} \tilde{\chi}(t-s) \sigma_t \sigma_s$, we defined
$Z_{\rm 1RSB}(\vb{h}) = \sum_{\vb{\sigma}} \exp \left (-H_{{\rm eff}}^{\rm 1RSB} \right )$ and 
denoted $\left \langle \cdots \right \rangle_{\rm 1RSB}$ as the average with respect to 
the Boltzmann distribution for the Hamiltonian $H_{{\rm eff}}^{\rm 1RSB}$.

In the 1RSB framework, the ``RS'' quasi-static solution is a special solution for which 
$\tilde{q}^0(t-s) = \tilde{q}^1(t-s) = \tilde{q}$ and $q^0(t-s) = q^1(t-s) = q$ hold. 
For examining the local stability of this solution against perturbations in the 1RSB direction, 
we set $\tilde{q}^1(t-s) = \tilde{q} + \Delta \tilde{q}(t-s)$
and $q^1(t-s) = q + \Delta q(t-s)$, and linearlize 
(\ref{eq:1RSB3}) and (\ref{eq:1RSB5}) using (\ref{eq:1RSB6}), 
which yields equations identical to (\ref{eq:Linearlization1}) and (\ref{eq:Linearlization2}). 
By repeating the same argument as before, the linearized equations can be diagonalized by the basis $\mathcal U$. 
This indicates that the critical mode is the 
$0$-th mode $({\mathcal U}_{t,0}) = (M^{-1/2},\ldots, M^{-1/2})^\top \in \mathbb{R}^M$, which is uniform over 
imaginary time, and 
the RS quasi-static solution is stable as long as (\ref{eq:AT_RS})  holds. 
This corresponds to 
the de Almeida--Thouless condition in classical spin glasses \cite{de1978stability}. 

\textcolor{black}{Our results partially support the treatment of \cite{PhysRevE.96.032112,PhysRevB.109.024431}, 
where the RSB for the quantum SK model is argued under the quasi-static ansatz. 
It may be reasonable 
on physical grounds that $q_{t,s}^{\mu,\nu}$
does not exhibit any imaginary-time dependence. 
If $q_{t,s}^{\mu,\nu}$
were to depend on the imaginary time variables
$t$ and $s$, the two replicas 
$\mu$ and $\nu$
would necessarily correspond to distinct equilibrium states carrying a nonzero spin current.
However, at least under the RS assumption, 
all replicas correspond to the same equilibrium state, and hence such a scenario is ruled out.}
\textcolor{black}{At the same time, however, the condition in (\ref{eq:general_stability}) suggests that the quasi-static approach may break down if $G^{\prime\prime}(x)$ exhibits decreasing regions for 
$x>0$. Investigating such scenarios would be an interesting direction for future research.}

\section{Numerical evaluation}
To examine the validity of the obtained stability condition for a quasi-static solution, 
we evaluated $\Gamma_{\rm c}$ numerically for the SK and Hopfield models. In the following, we focus on the cases of $h_0=0$.\par
Several earlier studies estimated $\Gamma_\mathrm{c}$ for the SK model using different methods (Table \ref{table1}). 
Refs. \cite{T_Yamamoto_1987,PhysRevB.76.184422} 
evaluated $\Gamma_\mathrm{c}$ utilizing  perturbation methods. 
By contrast, \cite{PhysRevE.96.032112,PhysRevB.109.024431} resorted to 
Monte Carlo methods. 
However, none of these methods directly satisfies the zero-temperature limit, 
and therefore, some form of extrapolation was performed.
We propose a method for estimating $\Gamma_\mathrm{c}$ that is different from previous approaches. It involves the following procedure.

\begin{figure}[htpb]
    \begin{tabular}{ccc}
    \begin{minipage}[t]{0.48\hsize}
    \centering
        \includegraphics[scale=0.6]{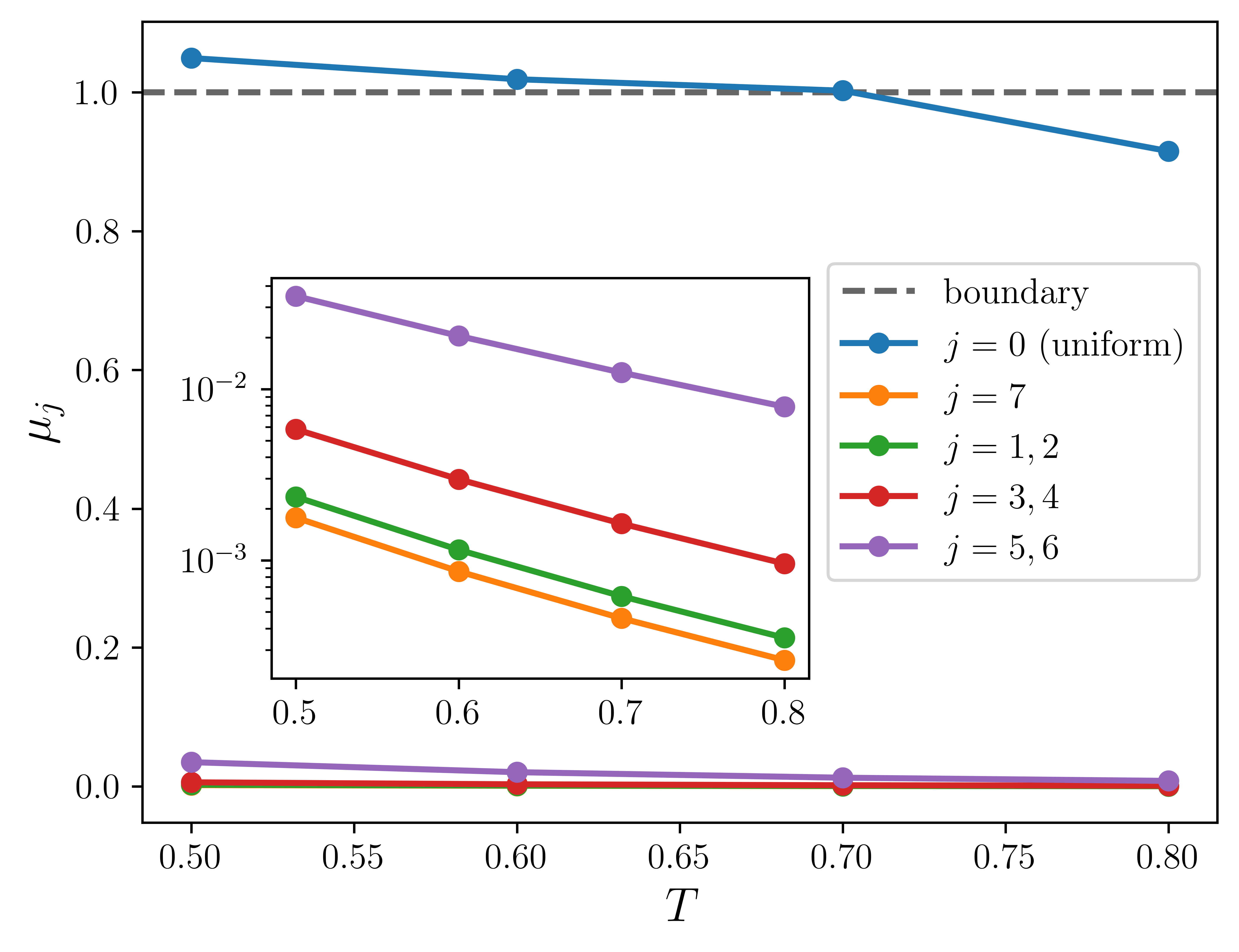}
        \caption{
        Stability coefficient 
        $\mu_j = 2\beta^2 G^{\prime \prime}
        \left (\beta \eta_j/M \right )
        \int {\mathrm{D}}z\left [T_j(z)\right ]^2$
        for each mode $j$ in the SK model with $M=8$, $h_0=0$, and $\Gamma =1.0$. 
        The mode of $j=0$ exceeds unity at a temperature
        between $T=0.7$ and $T=0.8$, which 
        indicates that the RS quasi-static solution becomes unstable below the temperature. 
         }\label{fig-1}
    \end{minipage}
    \begin{minipage}[t]{0.02\hsize}
        ~
    \end{minipage}
    \begin{minipage}[t]{0.48\hsize}
    \centering
        \includegraphics[scale=0.6]{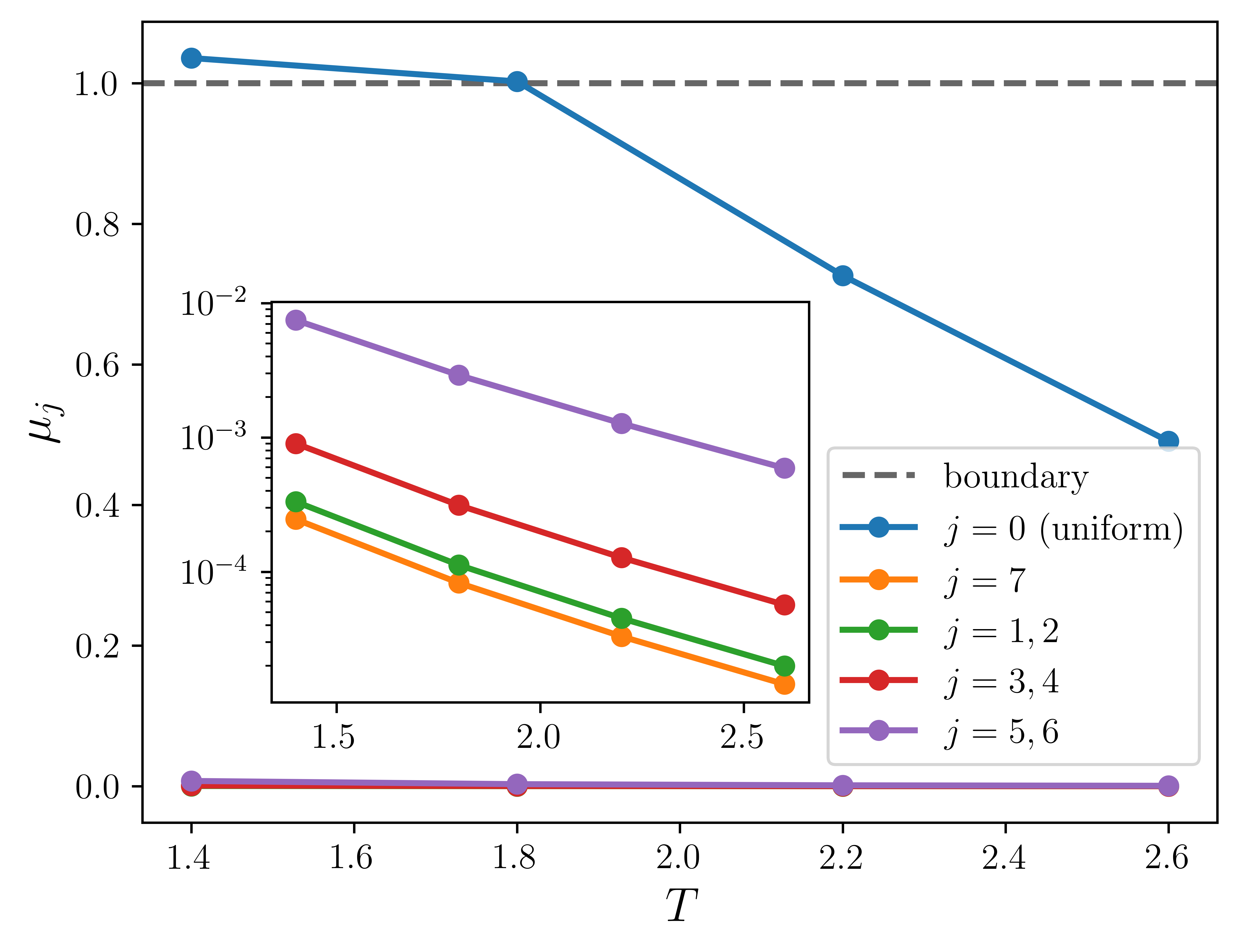}
        \caption{
            Stability coefficient 
        in the Hopfield model with $M=8$, $\alpha = 2.0$, $h_0=0$, and $\Gamma =1.0$. 
		As in the case of the SK model, that of $j=0$ exceeds unity 
		at a temperature between $T=1.8$ and $T=2.2$. 
            }\label{fig0}
    \end{minipage}
    \end{tabular}
\end{figure}

First, for a fixed finite Trotter number, $M$, 
the stability coefficient $\mu_j =2\beta^2 G^{\prime\prime}\left ( \frac{\beta \eta_j}{M} \right ) 
    \int {\mathrm D}z \left [ T_j(z) \right ]^2 $ in (\ref{eq:general_stability})
was evaluated for all $j \in \{0,1,\ldots, M-1\}$. 
We fixed the transverse field $\Gamma$ at $M=8$ and varied the temperature $T$. 
The behavior of $\mu_j$ is shown in Fig.\ref{fig-1} and Fig.\ref{fig0}.

For both models, the maximum value was consistently achieved by $\mu_0$, 
which corresponds to a uniform mode in imaginary time. 
This behavior does not depend on the Trotter number $M$, 
providing a numerical confirmation of the theoretical predictions presented in the previous section.  
Accordingly, we determined the critical temperature $T_\mathrm{c}$ 
as a function of the transverse field $\Gamma$ by identifying the point at which $\mu_0 = 1$, which represents the critical condition in  (\ref{eq:general_stability}), 
for various values of $M$. The results are shown in Fig. ~\ref{fig1} 
and Fig.~\ref{fig3}.

Next, we extrapolated the inflection points of 
$T_\mathrm{c}(\Gamma)$ as a function of the Trotter number $M$. This was motivated by the following considerations. 
The critical temperature $T_\mathrm{c}(\Gamma)$ of the quantum SK model
is expected to be convex upwards \cite{T_Yamamoto_1987,PhysRevE.96.032112,PhysRevB.76.184422,PhysRevLett.127.207204}. 
However, as shown in Fig. ~\ref{fig1} and 
Fig.~\ref{fig3}, $T_\mathrm{c}(\Gamma)$ of the finite 
Trotter number $M$ is not convex upwards. 
To recover the expected convex shape at the quantum limit $M\to \infty$, 
it is natural to assume that the inflection point of $T_\mathrm{c}(\Gamma)$  shifts toward  
zero as $M$ increases. 

This method has certain advantages over existing methods. 
First, it does not require many Trotter slices, $M$. 
Conventional approaches must maintain a  very large $M$ (up to $M = 512$) to satisfy the condition $TM>1$, 
which ensures the validity of the Suzuki--Trotter decomposition. 
However, 
our method naturally and simultaneously approaches the $T\to0$ limit and $M\to\infty$
while satisfying the constraint. 
In practice, the experiments were performed using $M\leq 20$.
Moreover, keeping $M$ small allows an exact evaluation of the expectation values 
over the effective Hamiltonian in $T_j$ 
thereby avoiding the Monte Carlo errors inherent in conventional methods.

To investigate the effectiveness of the method, we conducted experiments on the SK 
model and compared it with the 
$\Gamma_\mathrm{c}$ obtained based on previous studies. 
Similar experiments were conducted using the Hopfield model.

\subsection{SK model}
\begin{figure}[htpb]
    \begin{tabular}{ccc}
    \begin{minipage}[t]{0.48\hsize}
    \centering
        \includegraphics[scale=0.6]{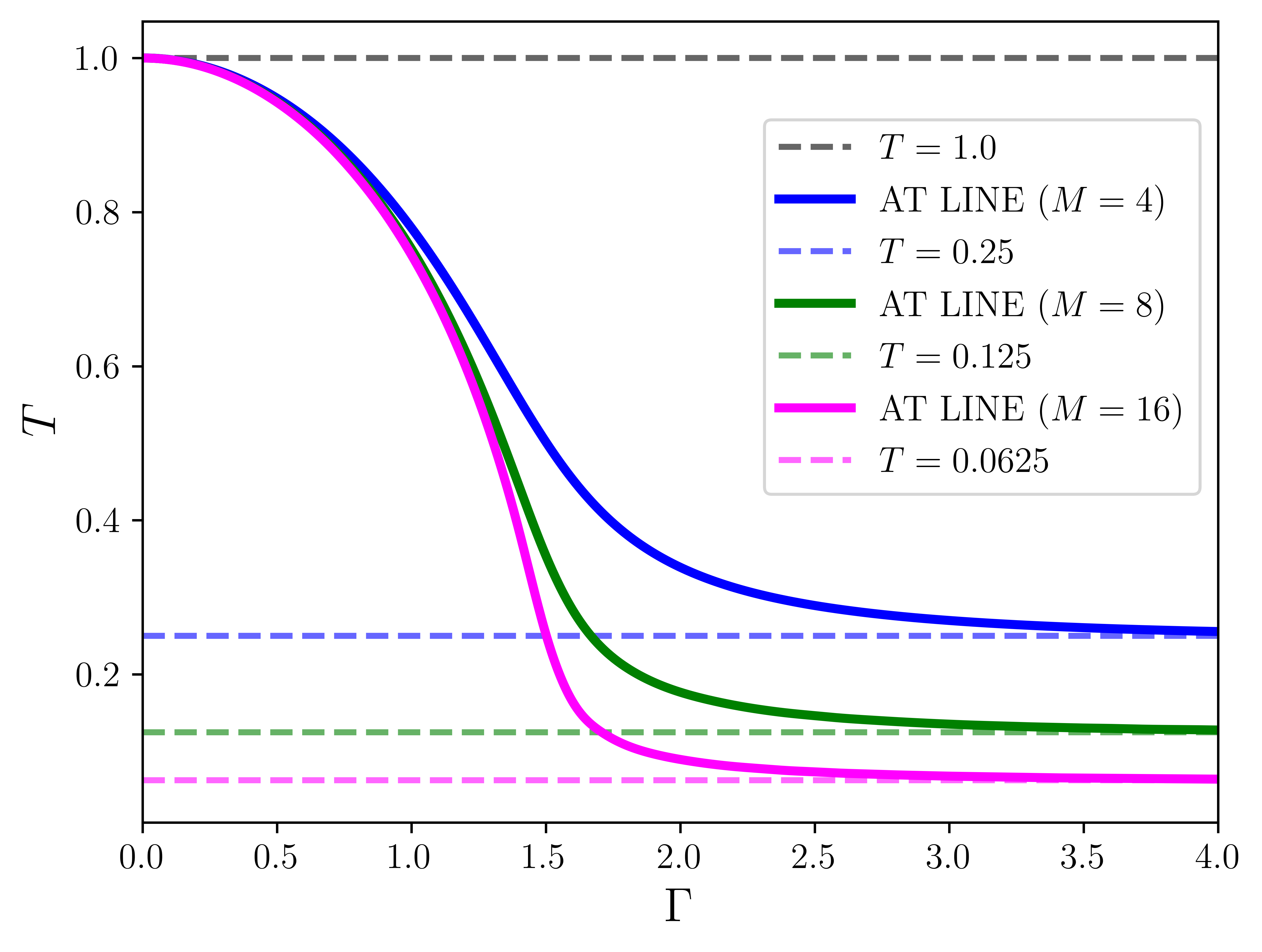}
        \caption{AT lines of the SK Model for $h_0 = 0$ are shown for $M=4$ (blue), $8$ (green), and $16$ (magenta). 
        The broken lines represent the analytical results in the limits $\Gamma \to 0$ (black) and 
        $\Gamma \to \infty$ (blue, green, and magenta). 
        }\label{fig1}
    \end{minipage}
    \begin{minipage}[t]{0.02\hsize}
        ~
    \end{minipage}
    \begin{minipage}[t]{0.48\hsize}
    \centering
        \includegraphics[scale=0.6]{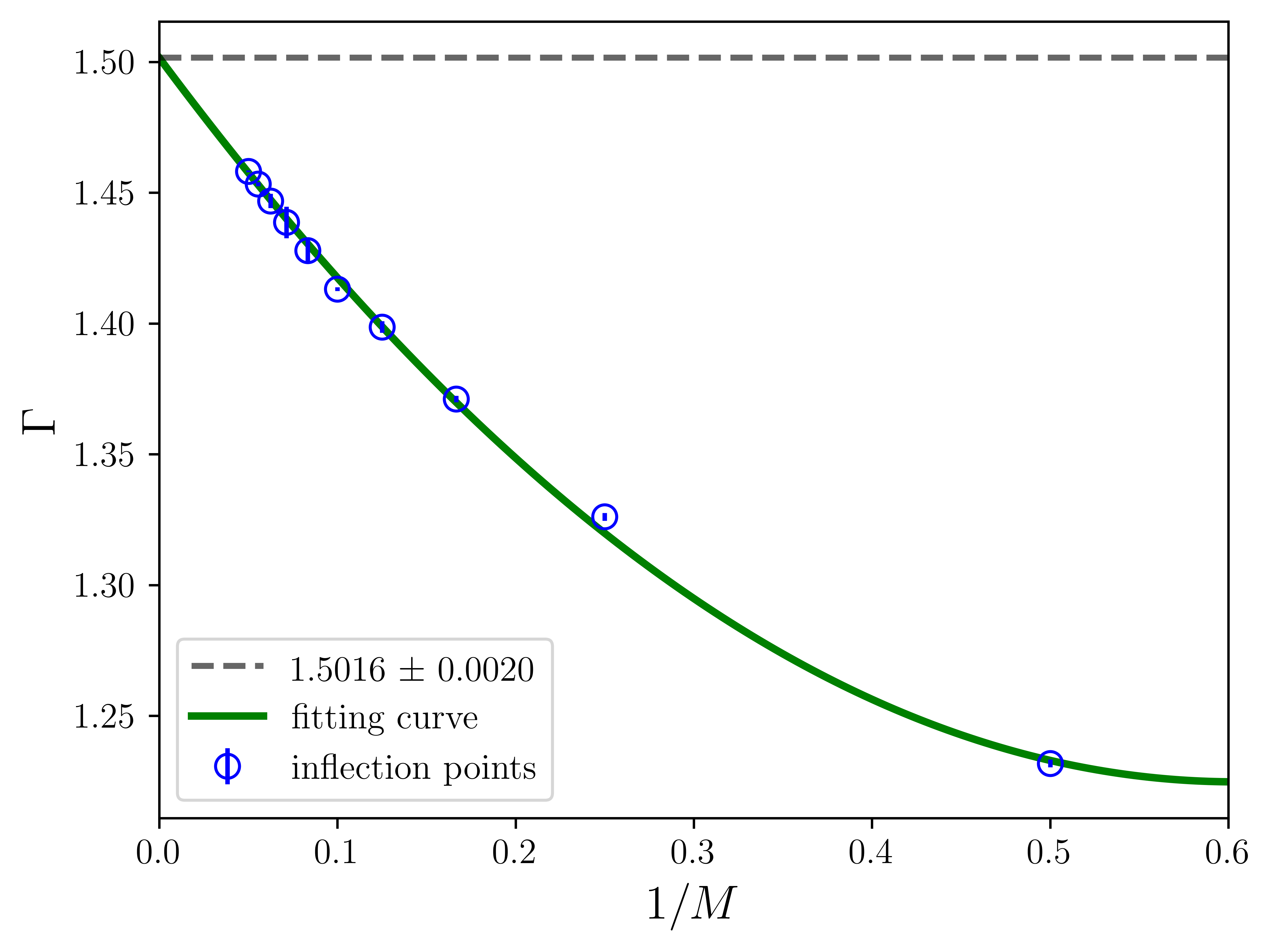}
        \caption{
            The inflection points in Fig. \ref{fig1} are plotted as a function of the finite Trotter number $M$. 
            These points are expected to converge to the critical value of the transverse field $\Gamma_\mathrm{c}$ at $T=0$ as $M\to\infty$. 
            A second-degree polynomial fit was performed with respect to $1/M$, and error bars were 
            estimated using the jackknife method, with the inflection points determined via interpolation based on the leave-one-out approach.
            }\label{fig2}
    \end{minipage}
    \end{tabular}
\end{figure}

Numerical experiments were conducted using the SK model for $M = 4, 8, 16$.  
The resulting boundaries between the RS and RSB phases are shown in Fig.\ref{fig1}. 
Since $\Gamma =0$ and $\Gamma = \infty$ 
correspond to two classical limits representing extremely strong and vanishing 
interactions between the Trotter replicas, the relationship $T'_\mathrm{c}=T_\mathrm{c}/M$ holds. 
This ensures that $T'_\mathrm{c}\to 0$ as $M\to\infty$, which is consistent with previous studies. 

Next, we investigated the behavior of the inflection point as a function of $M$, as shown in 
Fig.~\ref{fig2}. The data for $M\leq 20$ were extrapolated to the limit $M\to \infty$ 
using a second-degree polynomial fit in $1/M$, yielding 
\begin{align*}
    \Gamma_\mathrm{c} = 1.501 \pm 0.002.
\end{align*}
This result agrees well with those of earlier studies (Table \ref{table1}), demonstrating the validity of our method.
As expected, this value is significantly different from that obtained by the SA, $\Gamma_\mathrm{c}=2$.
\begin{table}[htb]
    \centering
    \caption{
    Estimates of the critical magnetic field
    $\Gamma_\mathrm{c}$ at the low-temperature limit
    for the SK model. 
    }
    \label{table1}
    \begin{tabular}{|c|c|c|} \hline
    Paper & $\Gamma_\mathrm{c}$ & Method\\\hline\hline
    -& 2 & SA\\\hline
    Yamamoto,Ishii,1987\cite{T_Yamamoto_1987} & 1.506 & Perturbation expansion\\\hline
    Takahashi,2007\cite{PhysRevB.76.184422} & 1.62 & Perturbation expansion\\\hline
    Young,2017\cite{PhysRevE.96.032112} & $1.51\pm 0.01$ & Suzuki Trotter($M\leq512$), $T$ fixed \\\hline
    Kiss,2024\cite{PhysRevB.109.024431} & 1.5 & QMC\\\hline
    Ours & $1.501\pm 0.002$ & Suzuki Trotter($M\leq 20$), $\Gamma$ fixed\\\hline
    \end{tabular}
\end{table}
\par

\subsection{Hopfield model}
\begin{figure}[htpb]
    \begin{tabular}{ccc}
    \begin{minipage}[t]{0.48\hsize}
    \centering
        \includegraphics[scale=0.6]{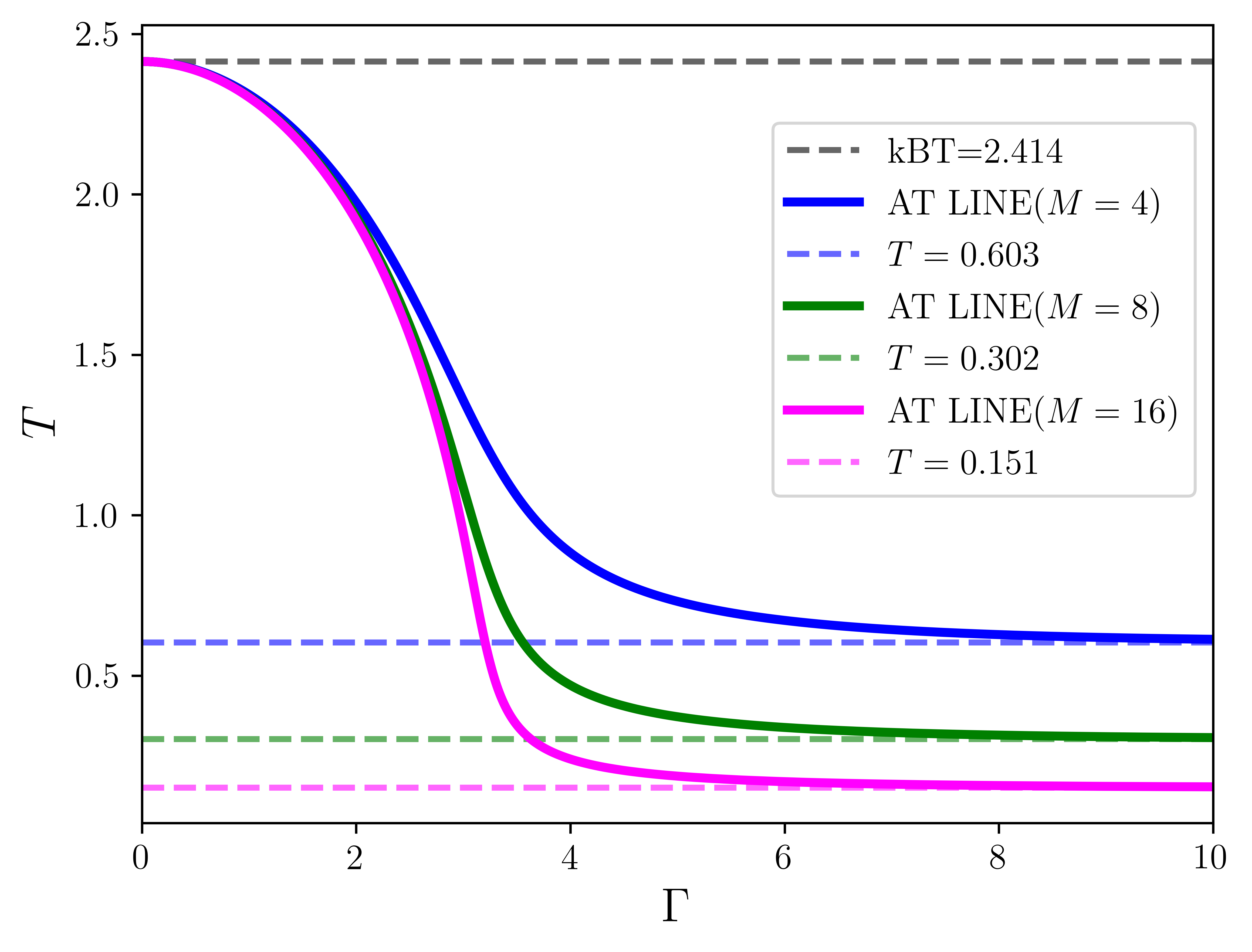}
        \caption{
        De Almeida--Thouless (AT) line of the Hopfield Model for $h_0 = 0$ and $\alpha = 2.0$, shown for $M = 4$ (blue), $8$ (green), $16$ (magenta). 
        The broken lines represent the analytical results  in the limits $\Gamma \to 0$ (black) and 
        $\Gamma \to \infty$ (blue, green, and magenta). }\label{fig3}
    \end{minipage}
    \begin{minipage}[t]{0.02\hsize}
        ~
    \end{minipage}
    \begin{minipage}[t]{0.48\hsize}
    \centering
        \includegraphics[scale=0.6]{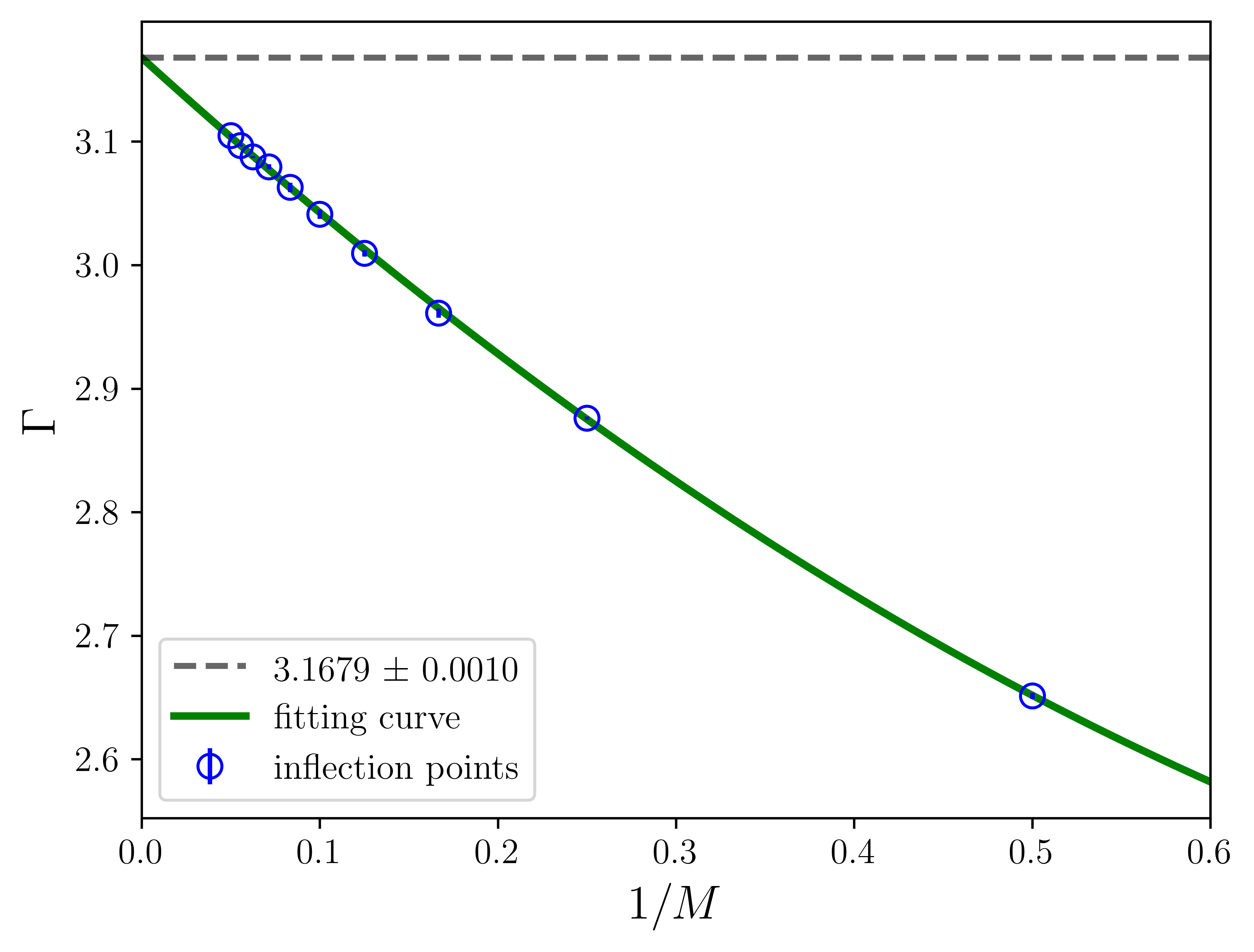}
        \caption{
	The inflection points in Fig. \ref{fig3} are plotted as a function of the finite Trotter number $M$. 
	As in Fig. \ref{fig2}, 
	these points are expected to converge to the critical value of the transverse field $\Gamma_\mathrm{c}$ at $T=0$ as $M\to\infty$. 
	}\label{fig4}
    \end{minipage}
    \end{tabular}
    \end{figure}

We also performed numerical experiments using the Hopfield model with $\alpha=2$.
The phase boundaries between RS and RSB phases for $M=4,8,16$ are shown in Fig. ~\ref{fig3}. 
The critical value of the transverse field $\Gamma_\mathrm{c}$ in the limit $T\to 0$
was estimated using the same procedure as that used for the SK model (Fig. ~\ref{fig4}). 
The estimated value of $\Gamma_\mathrm{c}$ converges to
\begin{align*}
    \Gamma_\mathrm{c} = 3.168 \pm 0.001
\end{align*}
which is also different from the results obtained by the SA.
\subsection{Random Orthogonal Model}
The two models discussed thus far undergo a continuous phase transition 
from the paramagnetic phase to the spin glass phase as determined by  (\ref{eq:general_stability}). This was derived by examining the effects of infinitesimal perturbations on the order parameters. 
By contrast, 
as mentioned in Section 2, the system in (\ref{eq:RIM}), 
\textcolor{black}{which is termed the ``random orthogonal model (ROM)'' 
\cite{Parisi_1995}}, exhibits 
RFOT, for which the critical condition cannot be characterized by 
the perturbative approach.  

\textcolor{black}{In the case of the ROM, a 1RSB solution characterized by  $q^1>0$ and $q^0 = 0$ emerges for  $m=1$ abruptly at a certain temperature $T_{\mathrm{D}}$, as a thermodynamically subdominant solution. Since positive overlap $q^1>0$ appears discontinuously from the RS solution of $q=0$, this scenario is referred to as a “random first-order transition (RFOT)”. Although this solution does not correspond to the equilibrium state, it is associated with an exponentially large number of thermodynamically isolated local minima, which makes equilibration extremely difficult. }

For the solution of this type,  (\ref{eq:1RSB1})--(\ref{eq:1RSB6}) are reduced to the following forms under qSA:
\begin{align}
    \tilde{\chi}(t-s) &= 2M\beta \sum_{j}{\mathcal U}_{t,j}{\mathcal U}_{s,j}G^\prime
    \left (\frac{\beta \eta_j}{M} \right ), \label{eq:ROM1RSB1} \\
    \tilde{q}^1
    &= \frac{2\beta }{m} 
    \left [
    G^\prime\left( \frac{\beta}{M} [\eta_0 + mMq^1] \right )
    -G^\prime\left( \frac{\beta \eta_0}{M}\right ) \right ], \label{eq:ROM1RSB2} \\
    \tilde{q}^0 &= 0, \label{eq:ROM1RSB3} \\
    \chi(t-s) &= 
    \frac{\int {\mathrm D} {y} Z_{\rm 1RSB}({h})^m 
    \left ( \left \langle \sigma_t \sigma_s \right \rangle_{\rm 1RSB}
    -\left \langle \sigma_t \right \rangle_{\rm 1RSB} \left \langle \sigma_s \right \rangle_{\rm 1RSB} \right )}
    {\int {\mathrm D} {y} Z_{\rm 1RSB}({h})^m},
    \label{eq:ROM1RSB4} \\
    q^1 &= 
    \frac{\int {\mathrm D}{y} Z_{\rm 1RSB}({h})^m 
    \left \langle \sigma_t \right \rangle_{\rm 1RSB} \left \langle \sigma_s \right \rangle_{\rm 1RSB} }
    {\int {\mathrm D}{y} Z_{\rm 1RSB}({h})^m}.  \label{eq:ROM1RSB5}\\
    q^0 &= 0. \label{eq:ROM1RSB6}   
\end{align}
Here, 
${h}=\sqrt{\tilde{q}^1}{y}$, and for $H_{{\rm eff}}^{\rm 1RSB}= -B \sum_{t}\sigma_t \sigma_{t+1} -M^{-1} \sum_{t} (\beta h_0 + h)\sigma_t 
-(2M^2)^{-1} \sum_{t,s} \tilde{\chi}(t-s) \sigma_t \sigma_s$, we defined
$Z_{\rm 1RSB}(\vb{h}) = \sum_{\vb{\sigma}} \exp \left (-H_{{\rm eff}}^{\rm 1RSB} \right )$ and 
denoted $\left \langle \cdots \right \rangle_{\rm 1RSB}$ as the average with respect to 
the Boltzmann distribution for the Hamiltonian $H_{{\rm eff}}^{\rm 1RSB}$.

\begin{figure}[htpb]
    \begin{tabular}{ccc}
    \begin{minipage}[t]{0.48\hsize}
    \centering
        \includegraphics[scale=0.6]{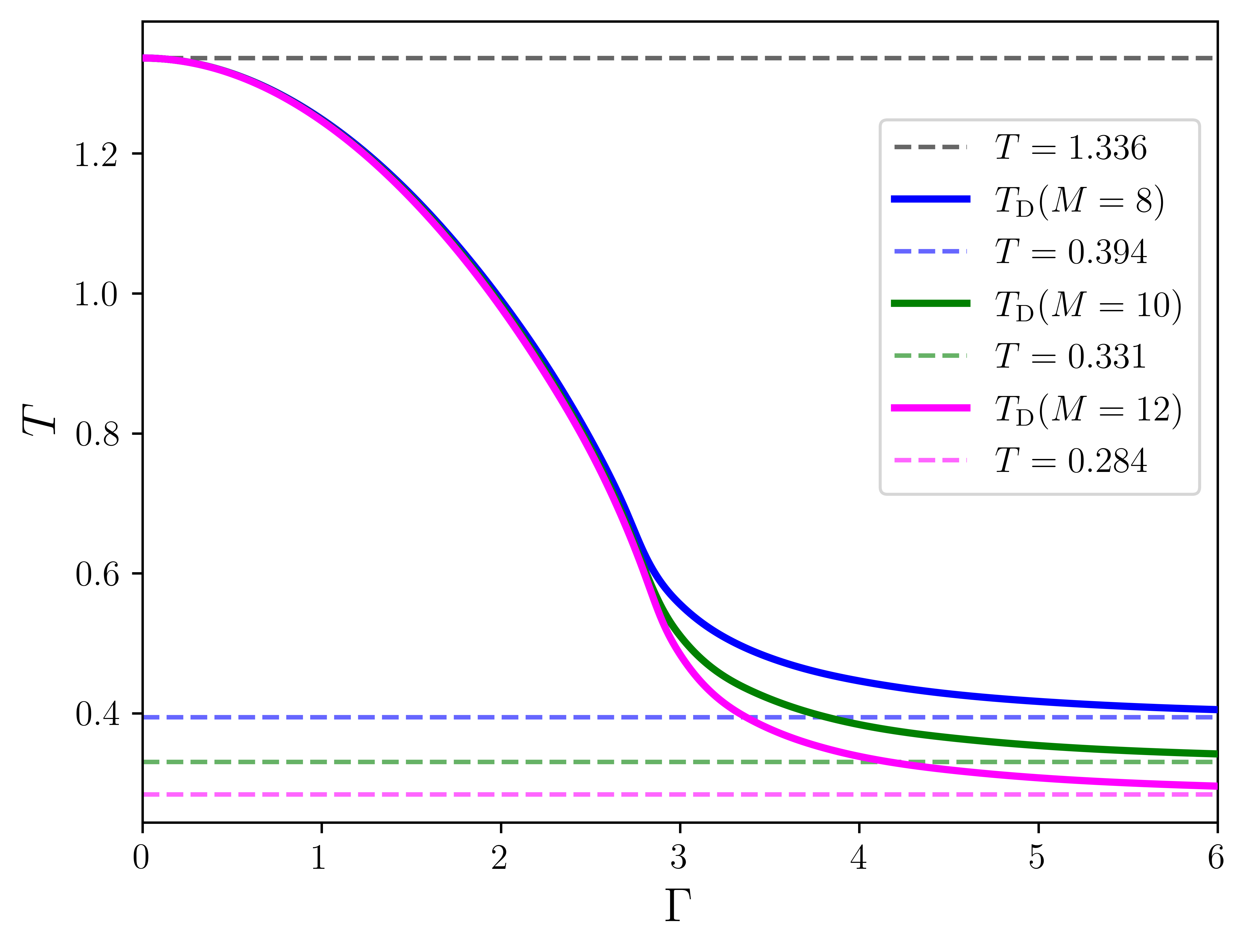}
        \caption{Critical temperatures $T_\mathrm{D}$ are plotted for finite Trotter size $M=8$ (blue), 10 (green), and 12 (magenta). 
        The broken lines represent analytical results in the limit $\Gamma\to \infty$.}\label{fig7}
    \end{minipage}
    \begin{minipage}[t]{0.02\hsize}
        ~
    \end{minipage}
    \begin{minipage}[t]{0.48\hsize}
    \centering
        \includegraphics[scale=0.6]{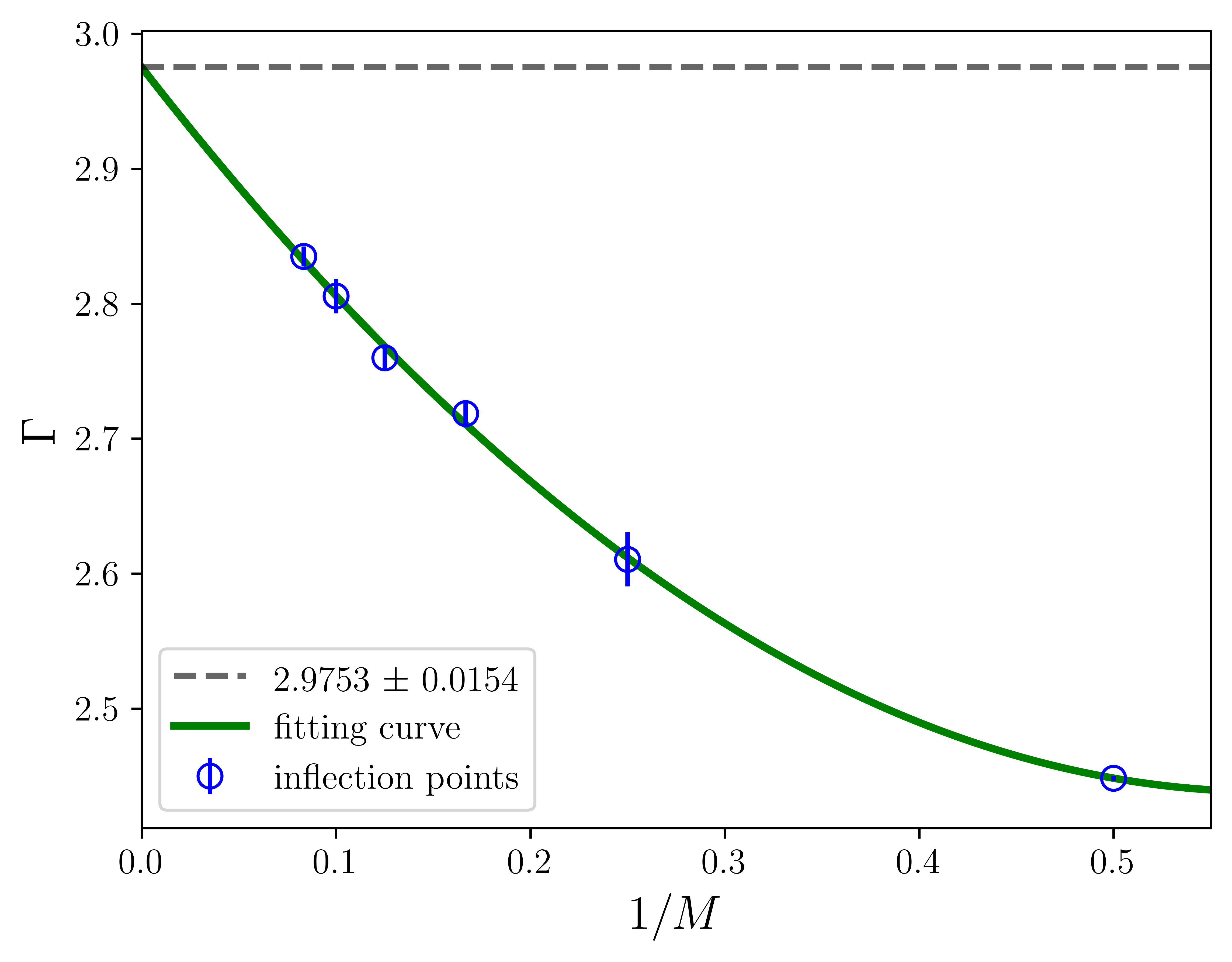}
        \caption{
        As in the cases of the SK and Hopfield models, we analyzed the inflection point of the phase boundary as a function of $1/M$, 
        and extrapolated the corresponding values of $\Gamma$ using a second-degree polynomial fit to $M\to \infty$. 
        The method used to estimate the error bars for each $M$ is the same as that employed for the SK and Hopfield models.
        }\label{fig8}
    \end{minipage}
    \end{tabular}
    \end{figure}

The dynamical transition temperature $T_\mathrm{D}$ is the temperature at which the 1RSB solution of $q^1>0$ 
appears to be a quasi-stable solution for $m=1$ 
whereas the critical temperature $T_\mathrm{c}$ is defined as  
the temperature at which the 1RSB solution becomes thermodynamically dominant. 
This is characterized by the condition $\Sigma(q^1;m) = 0$, where
\begin{align}
    \Sigma(q^1;m) &=m^2\pdv{\beta\phi}{m}\\
    &=\frac{m^2}{2}q^1\tilde{q}^1+G\qty(\frac{\beta}{M}\qty(mMq^1+\sum_{t}\chi_{t,s}))-m\beta q^1G'\qty(\frac{\beta}{M}(mMq^1+\sum_{t}\chi_{t,s} ))\\
   &\hspace*{5em}-G\qty(\frac{\beta}{M}\sum_{t}\chi_{t,s})+\ln\int\mathrm{D}y Z_{\rm 1RSB}({h})^m -m\frac{\displaystyle\int\mathrm{D}y (\ln Z_{\rm 1RSB}({h}))Z_{\rm 1RSB}({h})^m}{\displaystyle\int\mathrm{D}y Z_{\rm 1RSB}({h})^m}
\end{align}
is termed ``complexity,'' representing the entropy of metastable states. 
In the classical limit, $\Gamma\to 0$, the transition temperatures are given by
\begin{align}
\label{eq:ROM_classical1}
    T_{\mathrm{D}}=0.134J\\
    \label{eq:ROM_classical2}
    T_{\mathrm{c}}=0.065J
\end{align}
as reported by \cite{Parisi_1995}. Here, we set $J=10$ and investigated the behavior of $T_\mathrm{D}$  
as a function of the transverse field $\Gamma$ for $M=8,10,12$. The results are presented in Fig. \ref{fig7}.

\begin{figure}[htpb]
\begin{tabular}{ccc}
\begin{minipage}[t]{0.48\hsize}
\centering
\includegraphics[scale=0.6]{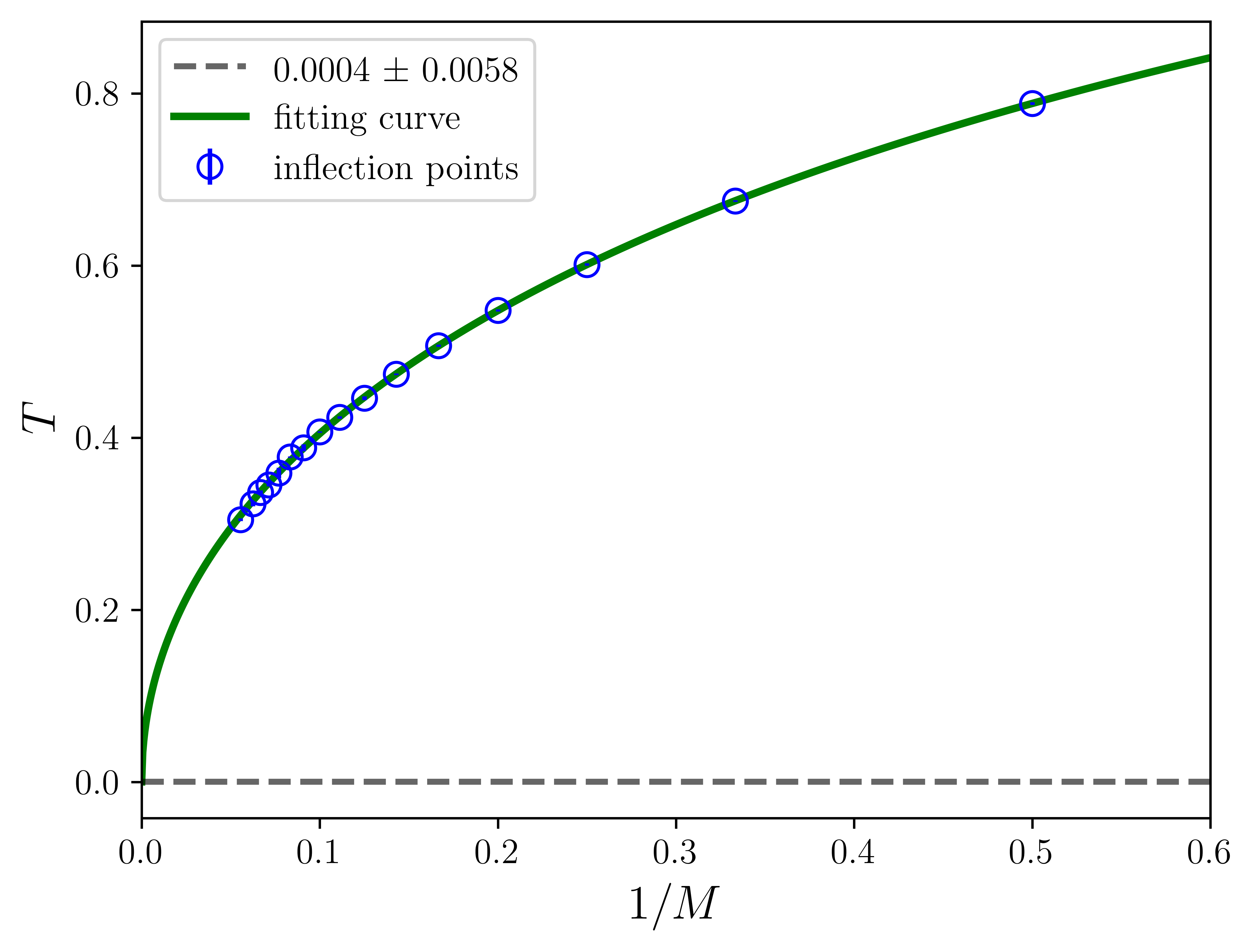}
\caption{
For the SK model, the temperatures $T$ at the inflection points are
plotted as a function of $M$. 
These values are extrapolated to the $M\to \infty$ limit using a second-degree polynomial fit with respect to $1/\sqrt{M}$, which minimizes the residual error among several methods. 
The extrapolation
yields a value that is nearly zero. A similar result is also obtained for the Hopfield model. 
}\label{fig-2}
\end{minipage}
\begin{minipage}[t]{0.02\hsize}
    ~
\end{minipage}
\begin{minipage}[t]{0.48\hsize}
\centering
\includegraphics[scale=0.6]{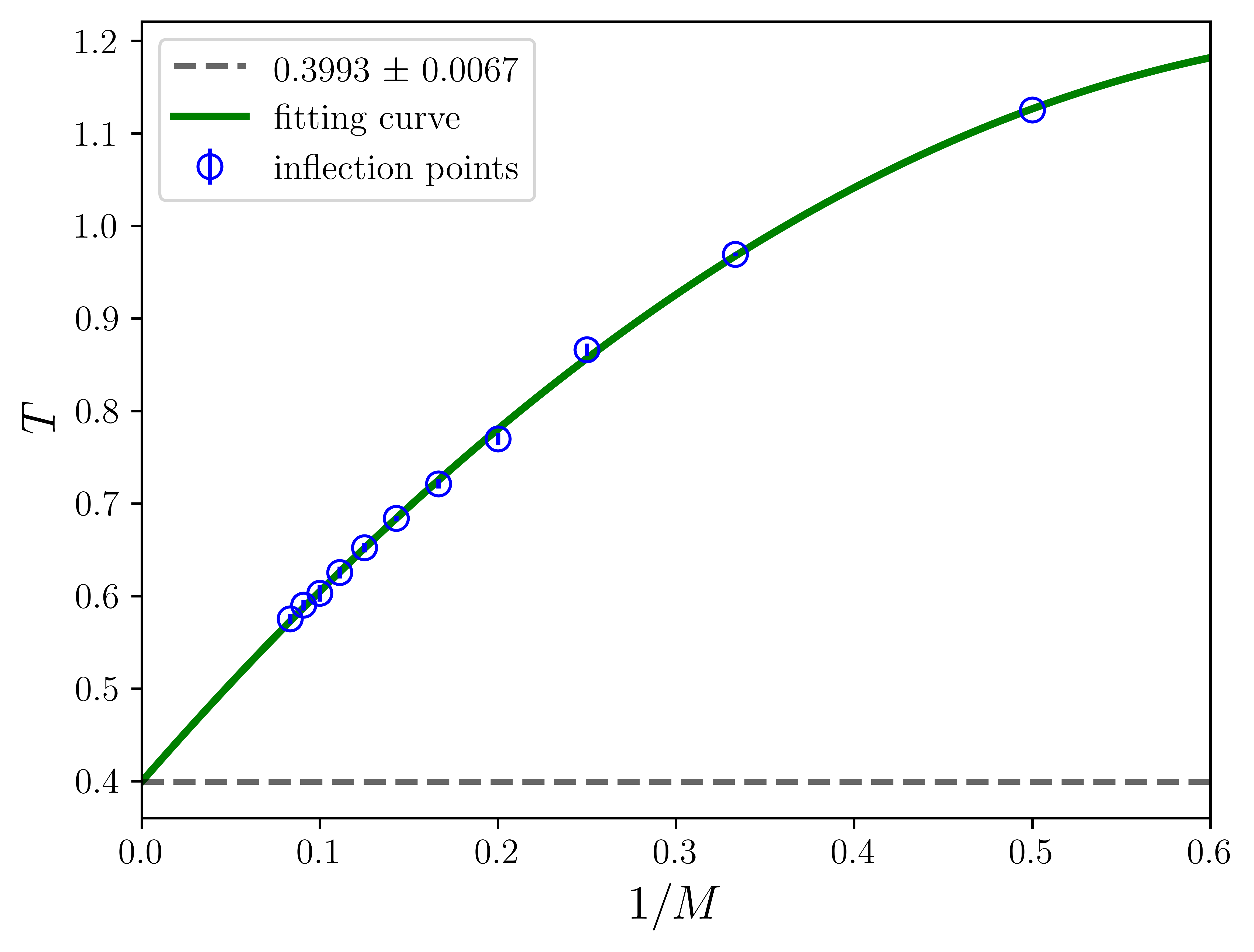}
    \caption{
    For the ROM, the temperatures $T$ at the inflection points are plotted as a function of $M$. 
    These values are extrapolated to the $M\to \infty$ limit using a second-degree polynomial fit with respect to $1/{M}$, 
    which minimizes the residual error among several methods. 
    In contrast to the SK model, a significantly large value remains in the limit. 
    }\label{fig-3}
\end{minipage}
\end{tabular}
\end{figure}

As in the previous two models, an inflection point appears 
in $T_\mathrm{D}(\Gamma)$. In Fig. \ref{fig8}, 
we applied a second-degree polynomial 
fit of the inflection point with respect to $1/M$, 
which estimates 
\begin{align*}
    \Gamma_\mathrm{I} = 2.97 \pm 0.01
\end{align*}
for the quantum limit $M\to \infty$. 
However, unlike in the SK and Hopfield models, it remains unclear whether the phase boundary 
in this model is also convex upward.
Consequently, it is uncertain whether the observed value corresponds to the critical transverse field in the low-temperature limit. 
The extrapolated temperature at the inflection point in the SK and Hopfield models 
converges to a value close to zero when $M \to \infty$ (Fig. \ref{fig-2}).
By contrast, for the ROM, a significantly large finite value persists, 
as shown in Fig. \ref{fig-3}.

\begin{figure}[htpb]
\begin{tabular}{ccc}
\begin{minipage}[t]{0.48\hsize}
\centering
\includegraphics[scale=0.6]{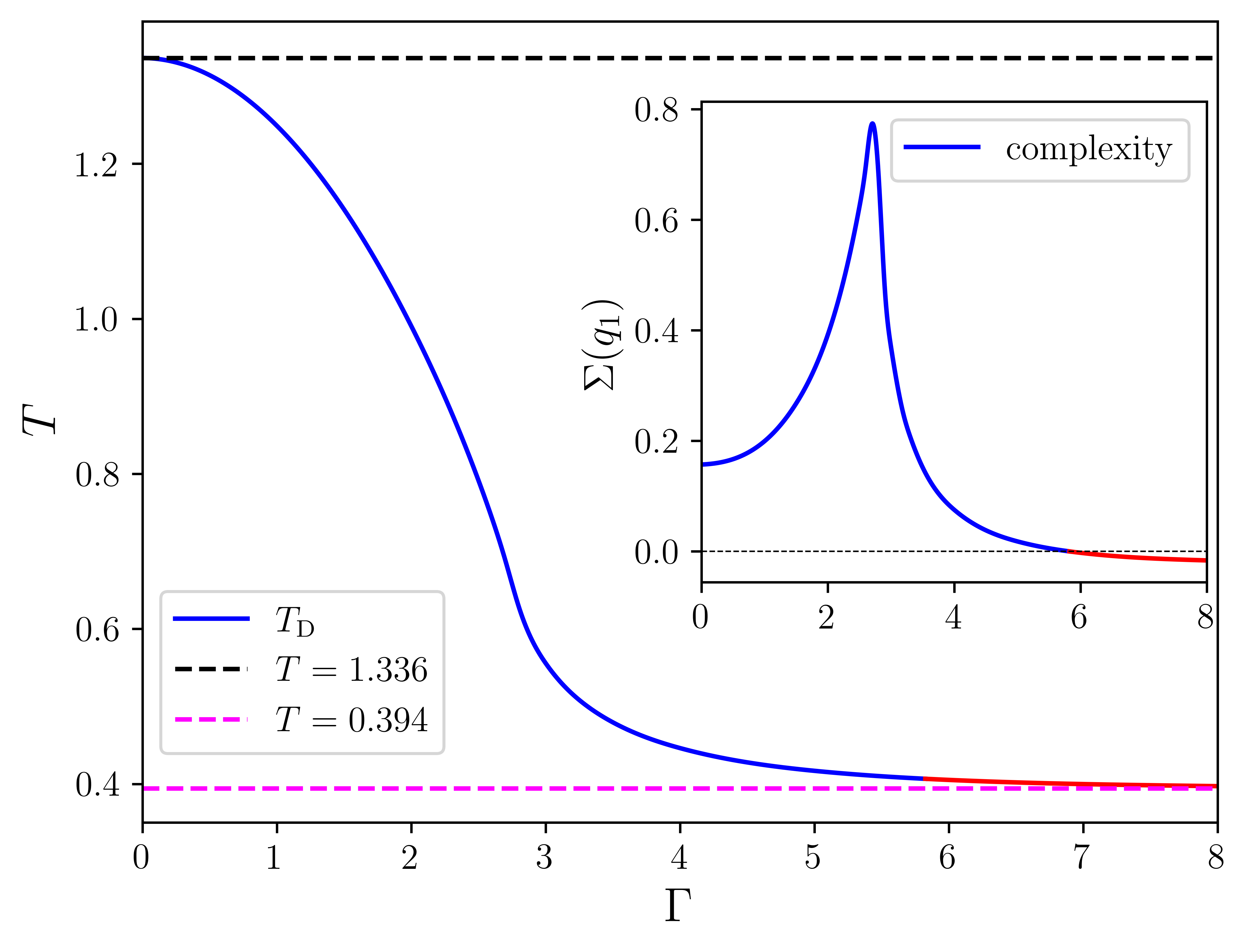}
\caption{
Dynamical transition temperature $T_\mathrm{D}$ is shown as a function of 
the transverse field $\Gamma$ for the ROM with $M=8$. 
As $\Gamma$ increases, the complexity at $T_\mathrm{D}$ becomes negative (inset). 
}\label{fig9}
\end{minipage}
\begin{minipage}[t]{0.02\hsize}
    ~
\end{minipage}
\begin{minipage}[t]{0.48\hsize}
\centering
\includegraphics[scale=0.6]{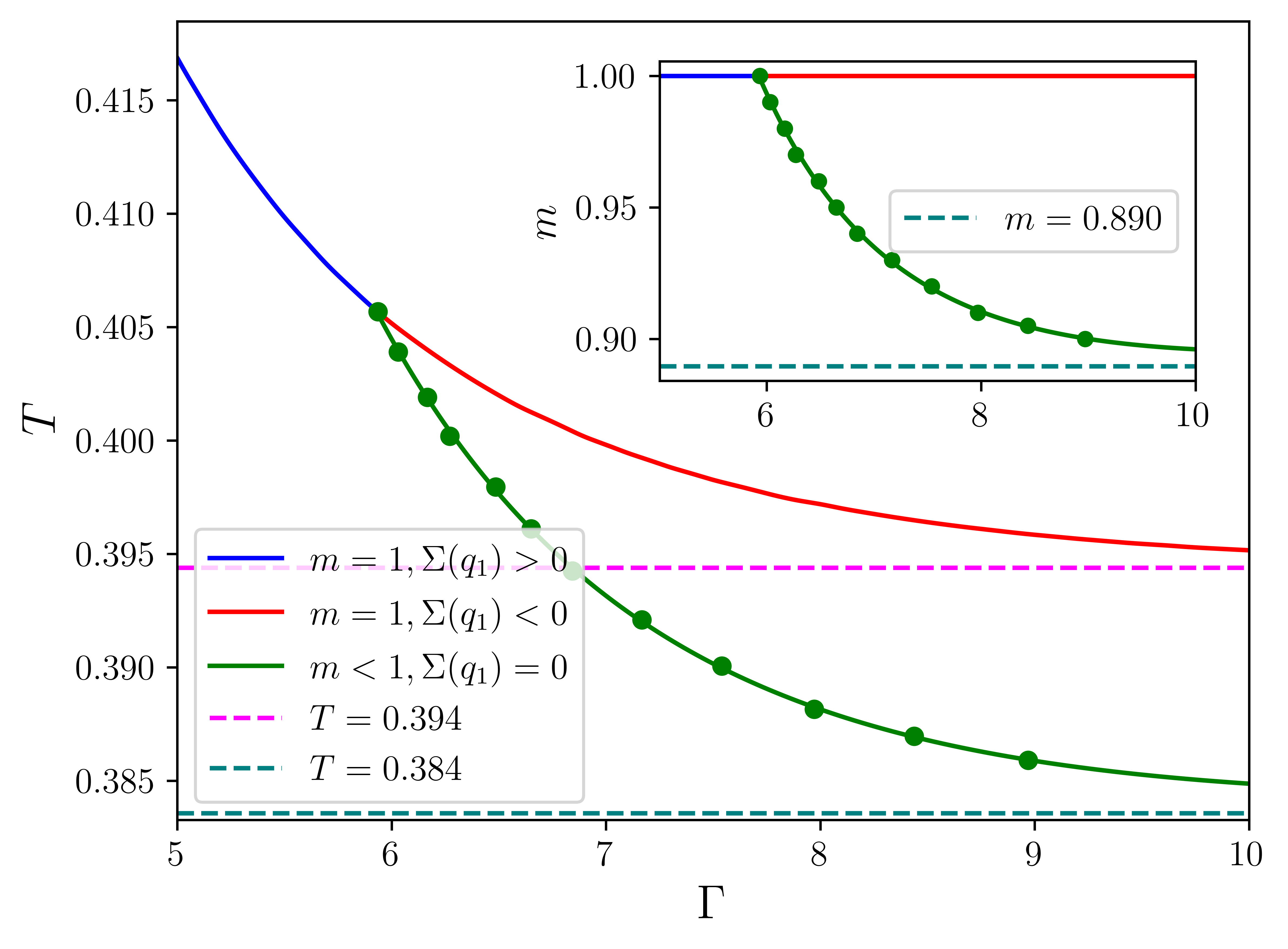}
    \caption{
    Phase diagram for the ROM with $M=8$ is shown. The blue curve represents the dynamical transition temperature $T_\mathrm{D}$. 
    As $\Gamma$ increases beyond a critical value $\Gamma_\mathrm{c}$, the complexity at $T_\mathrm{D}$ becomes negative (indicated in red), 
    which is physically unacceptable. 
    Therefore, for $\Gamma > \Gamma_\mathrm{c}$, the phase boundary (green) is determined by tuning $m\in [0,1]$ 
    so that the complexity vanishes. 
    }\label{fig10}
\end{minipage}
\end{tabular}
\end{figure}\par

A detailed investigation of the case $M=8$
suggests an anomalous behavior of the complexity at $T_\mathrm{D}$  
as plotted in Fig. \ref{fig9}.
Equations (\ref{eq:ROM_classical1}) and (\ref{eq:ROM_classical2}) indicate
that a temperature region exists in which $q^1>0$
and $\Sigma(q^1;m=1)>0$ as $\Gamma\to 0$. 
However, as $\Gamma$ increases from $0$, 
the region disappears when $\Gamma$ exceeds the critical value $\Gamma_\mathrm{c}=5.93$, 
indicating that $\Sigma(q^1;m=1)<0$ for $\Gamma > \Gamma_\mathrm{c}$. 

Consequently, $T_\mathrm{D}$ does not exist for $\Gamma>\Gamma_\mathrm{c}$. 
This implies that when $T$ is reduced from a sufficiently high value for fixed $\Gamma > \Gamma_\mathrm{c}$, the 1RSB solution for
$q^0 = 0$ and $q^1 > 0$ with $\Sigma(q^1;m)=0$ and $0<m<1$
suddenly appears, taking over the thermodynamic dominance
from the RS solution of $q=0$ at $T_\mathrm{c}$.

\begin{figure}[htpb]
    \begin{tabular}{ccc}
    \begin{minipage}[t]{0.48\hsize}
        \centering
        \includegraphics[scale=0.6]{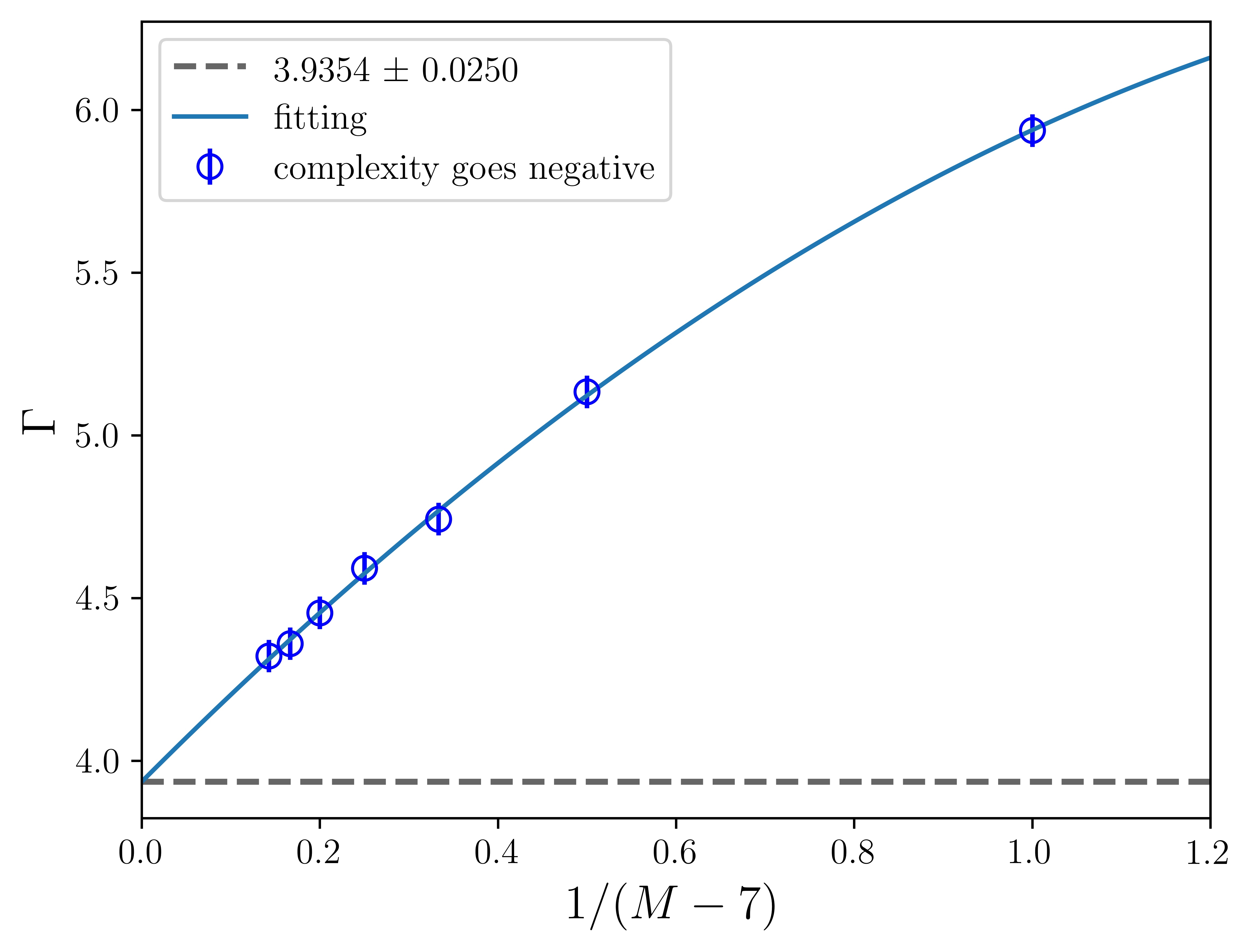}
        \caption{
	The values of $\Gamma$ at which the complexity vanishes at $T_\mathrm{D}$ were extrapolated to the $M\to \infty$ limit 
	using data for $M=8, 9, \ldots, 14$. Because the values of $\Gamma$ do not exist for $M\le 7$, the extrapolation was 
	performed by a second-degree polynomial fit with respect to $1/{(M-7)}$. 
        }\label{fig11}
    \end{minipage}&
    \begin{minipage}[t]{0.02\hsize}
        ~
    \end{minipage}&
    \begin{minipage}[t]{0.48\hsize}
    \centering
    \includegraphics[scale=0.6]{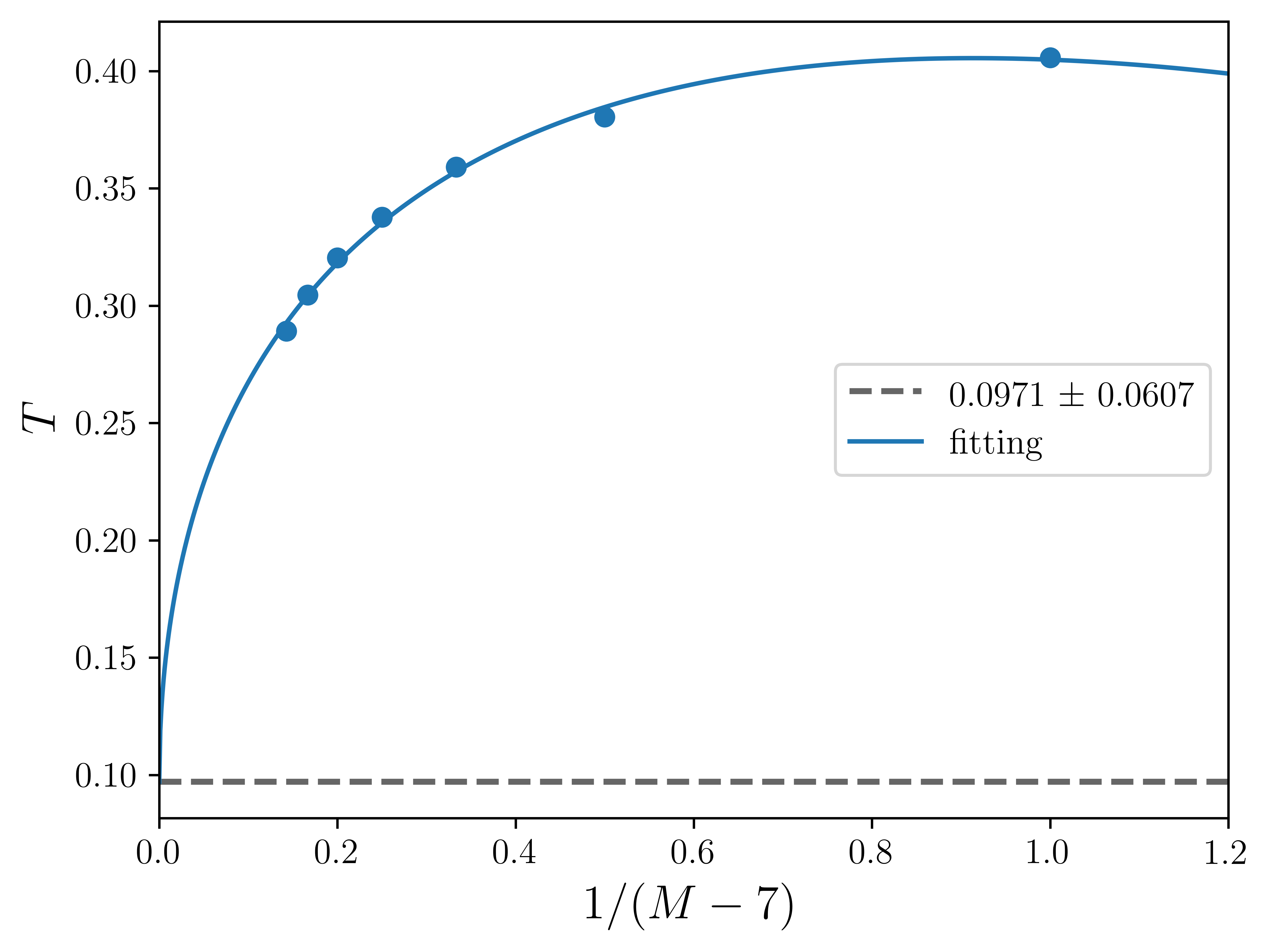}
        \caption{
        The values of $T$ at which the complexity vanishes at $T_\mathrm{D}$ were extrapolated to the $M\to\infty$ limit using the data for $M=8, 9, \ldots, 14$. The extrapolation was 
	performed using a second-degree polynomial fit with respect to $1/\sqrt{(M-7)}$, which minimizes the residual error 
	among several methods. The error is estimated using the leave-one-out jackknife method applied to seven data points. 
	Nonzero $\hat{T}_\mathrm{c}$ remains even in the limit $M\to\infty$.
        }\label{fig14}
    \end{minipage}
    \end{tabular}
    \end{figure}

Fig. \ref{fig11} shows the extrapolated curve for $\Gamma_\mathrm{c}$ obtained
using data from $M=8$ to $M=14$. 
Considering the limit of $M\to \infty$, 
this estimates 
\begin{align*}
    \Gamma_\mathrm{c} = 3.94 \pm 0.03
\end{align*}
for the quantum systems. 
This value is significantly different from that of $\Gamma_\mathrm{I}$. 
We also investigated the behavior of the critical temperature $T_\mathrm{D}$ at $\Gamma_\mathrm{c}$ in the limit $M \to \infty$ by extrapolating the data obtained for $M = 8$ to $M = 14$ (Fig. \ref{fig14}).
The results suggest that $T_\mathrm{D}$ 
remains finite even in the quantum limit, yielding
\begin{align} 
T_\mathrm{D}(\Gamma_\mathrm{c}) = 0.0971 \pm 0.0607.
\label{eq:TDTc_ROM}
\end{align}

\begin{figure}[htpb]
    \begin{tabular}{ccc}
    \begin{minipage}[t]{0.48\hsize}
    \centering
    \includegraphics[scale=0.6]{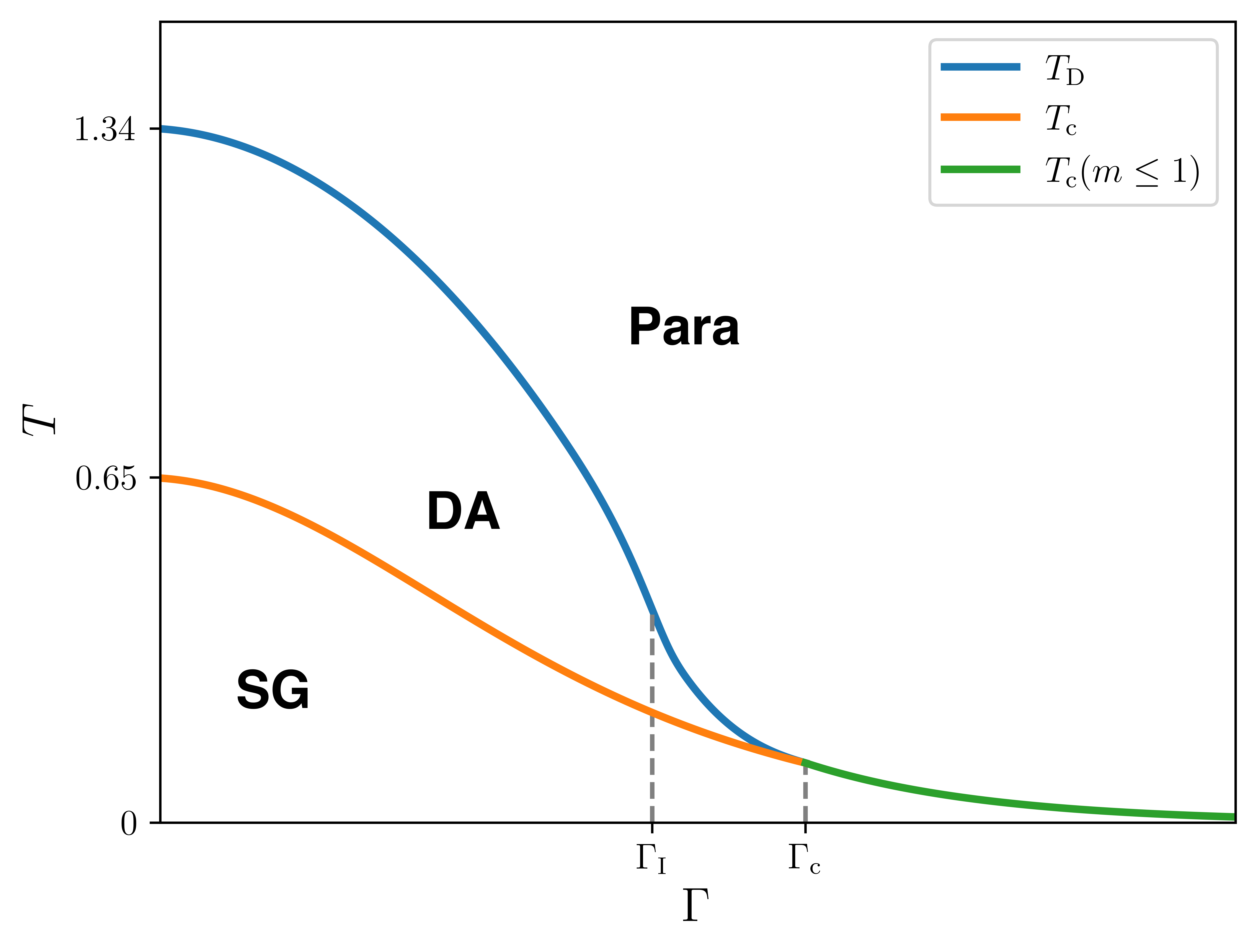}
        \subcaption{
        }\label{fig12}
    \end{minipage}
    
    \begin{minipage}[t]{0.48\hsize}
        \includegraphics[scale=0.6]{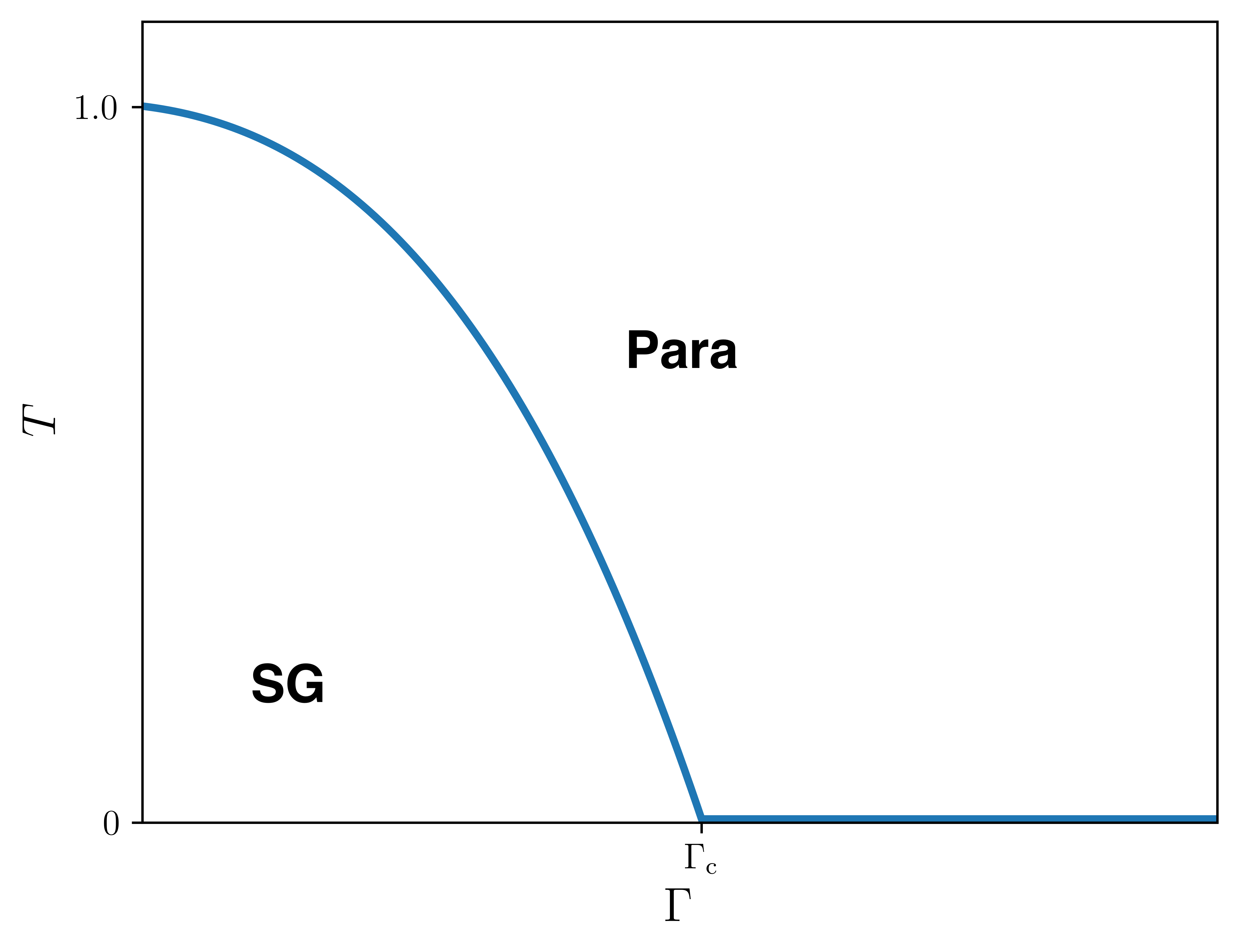}
        \subcaption{
        }\label{hoge}
    \end{minipage}&
    \begin{minipage}[t]{0.02\hsize}
        ~
    \end{minipage}&
    \end{tabular}
    \caption{
        Schematic representations of the expected phase diagrams 
        for the quantum (a) ROM and (b) SK model, conjectured in the limit $M\to \infty$. 
        }
    \end{figure}

Because a quantum system is expected to remain in a paramagnetic state within the limit 
$\Gamma \to \infty$ even as 
$T\to 0$, the above considerations suggest the schematic phase diagram of the ROM in the quantum limit, as shown in Fig. ~\ref{fig12}.
Although the finite value in (\ref{eq:TDTc_ROM}) may reflect the systematic 
error arising from the extrapolation, meaning that the true value could vanish, this nonetheless indicates that the effect of quantum noise in the ROM differs fundamentally from that of thermal noise. As illustrated in Fig.\ref{hoge}, the phase transition scenarios in 
the SK model are symmetric along both the temperature $T$ and transverse field 
$\Gamma$ axes. By contrast, in the ROM, the “dynamically arrested (DA)” phase disappears along the $\Gamma$ axis 
because of quantum effects. 
In classical systems, critical slowing down occurs during the DA phase, which makes equilibration extremely difficult~\cite{E_Marinari_1994}. 
The disappearance of the DA phase along the $\Gamma$ axis in the ROM may therefore indicate that 
quantum noise is more effective than thermal noise in facilitating equilibration.

\section{Summary and discussion}
In this study, we investigated the imaginary-time dependence of the order parameter in the transverse-field Ising model with rotationally invariant random interactions using the Suzuki--Trotter decomposition and the replica method. 

Unlike previous approaches that rely on extensive Monte Carlo simulations 
combined with the Suzuki-Trotter formula \cite{PhysRevE.96.032112} or the quantum Monte Carlo method \cite{PhysRevB.109.024431}, our method estimates the quantum limit behavior 
by numerically extrapolating the exact results obtained from systems with 
a small number of Trotter slices $M$. 
For the SK model, our approach yields a critical value of the transverse field for the zero-temperature 
QSG transition, 
$\Gamma_\mathrm{c}$, that closely agrees with earlier estimates  
for systems with $M\le 20$.
For the Hopfield model, we provide what appears to be the first estimate of 
$\Gamma_\mathrm{c}$.
For the random orthogonal model, our analysis suggests that the RFOT scenario is altered by quantum effects
in the vanishing temperature limit $T\to 0$. Unlike in the classical case, the thermodynamic dominance of the RS solution is directly overtaken by the 1RSB solution with vanishing complexity. 

\textcolor{black}{Our method is based on the idea of extrapolating the results obtained for systems with a small number of Trotter slices $M$, for which the free energy can be assessed “exactly” under the replica symmetric assumption, to the limit $M \to \infty$. This allowed us to evaluate the critical points of systems other than the SK model—cases that have not been explored in previous studies—with relative ease. However, we acknowledge that these results are not entirely conclusive, and they should be compared with results obtained by other methods such as quantum Monte Carlo simulations. That said, since such comparisons would require significantly more computational effort, we leave them as a subject for future work.}

We demonstrated that a sufficient condition for the spin-glass order parameter to be uniform in imaginary time across different replicas is in accordance with the AT stability condition, which governs the stability of the replica structure, at least within the RS phase. This finding supports the validity of the qSA, which is 
the approach proposed in \cite{PhysRevE.96.032112,PhysRevB.109.024431}, at least in the RS regime.
Another commonly used approximation, the SA \cite{bray1980replica}, assumes uniformity in imaginary time for the spin-glass order parameters within individual replicas. We demonstrated that the application of this approximation leads to inaccurate estimates of the critical transverse field for the QSG transition in the SK model, indicating that the approximation is not appropriate for this purpose.

When RFOT occurs in classical systems, it yields critical slowing down below the dynamical transition temperature $T_\mathrm{D}$, 
which hinders convergence to a thermodynamic equilibrium state.
Our results for the ROM suggest that quantum effects may help avoid 
this slowing down during the annealing process.
It would be interesting to explore these effects using a real quantum annealing device, such as the D-Wave machine \cite{johnson2011quantum}.

\begin{acknowledgments}
The authors gratefully acknowledge insightful discussions with Koki Okajima and Takashi Takahashi.
This work was supported by the Forefront Physics and Mathematics Program to Drive Transformation (FoPM), the World-leading Innovative Graduate Study (WINGS) Program, the University of Tokyo (YH), 
and MEXT/JSPS KAKENHI Grant Number 22H05117 (YK). 
\end{acknowledgments}

\end{document}